\documentstyle[12pt]{article}
\newcommand{\eq}{\begin{equation}}
\newcommand{\en}{\end{equation}}
\newcommand{\eqn}{\begin{eqnarray}}
\newcommand{\enn}{\end{eqnarray}}
\newcommand{\CR}{\nonumber \\}
\newcommand{\ad}{{\rm ad}}
\newcommand{\BC}{{\bf C}}
\newcommand{\bchi}{\bar{\chi}}
\newcommand{\bG}{\bar{G}}
\newcommand{\BG}{{\bf g}}
\newcommand{\Bn}{{\bf n}}
\newcommand{\Bh}{{\bf h}}
\newcommand{\cM}{{\cal M}}
\newcommand{\cF}{{\cal F}}
\newcommand{\cN}{{\cal N}}
\newcommand{\EXP}{{\rm exp}}
\newcommand{\E}{{\rm e}}
\newcommand{\I}{{\rm i}}
\newcommand{\half}{{1\over2}}
\newcommand{\HBG}{\hat{{\bf g}}}
\newcommand{\pa}{\partial}
\newcommand{\rank}{{\rm rank}}

\newcommand{\tr}{{\rm tr}}

\newcommand{\AT}[2]{{\scriptstyle #1}\atop{\scriptstyle #2}}
\newcommand{\A}{\alpha}
\newcommand{\B}{\beta}
\newcommand{\D}{\delta}
\newcommand{\DE}{\Delta}
\newcommand{\G}{\gamma}

\newcommand{\vep}{\varepsilon}

\newcommand{\T}{\theta}

\newcommand{\LM}{\Lambda}
\newcommand{\lm}{\lambda}

\newcommand{\vp}{\varphi}
\newcommand{\PL}{Phys. Lett.}
\newcommand{\CMP}{Commun. Math. Phys.}

\newcommand{\IJMP}{Int. J. Mod. Phys.}
\newcommand{\NPB}[1]{Nucl. Phys. {\bf B#1}}
\newcommand{\PLB}[1]{Phys. Lett. {\bf B#1}}
\newcommand{\MPLA}[1]{Mod. Phys. Lett. {\bf A#1}}
\newcommand{\IJMPA}[1]{Int. J. Mod. Phys. {\bf A#1}}

\renewcommand{\thesection}{\arabic{section}}

\newcommand{\sect}[1]{\section{#1}\setcounter{equation}{0}}
\newcommand{\subsect}[1]{\subsection{#1}}
\newcommand{\alphsec}{\setcounter{section}{0}
                      \renewcommand{\thesection}{\Alph{section}}}
\newcommand{\appe}[1]{\setcounter{equation}{0}
                      \setcounter{subsection}{0}
                      \addtocounter{section}{1}
                      \section*{Appendix \ #1}}
\def\bA{\bar{\alpha}}
\def\bLM{\bar{\Lambda}}
\def\blm{\bar{\lambda}}
\def\bPhi{\bar{\Phi}}
\newcommand{\dz}{{d z\over 2\pi\I}}
\def\ord#1#2{{#2\over (z-w)^{#1}}}
\def\ordo#1{{#1\over z-w}}

\def\sdim{{\rm sdim}}
\def\WG{\widehat{G}}
\def\hphi{\hat{\phi}}
\def\hJ{\hat{J}}

\def\ppsi{\pa\psi}
\def\Tb{\tilde{b}}
\def\Tc{\tilde{c}}
\def\TT{\tilde{T}}
\def\TG{\tilde{G}}
\newcommand{\R}{{ }^{r}\hspace{-3pt}}

\newcommand{\zwth}[1]{\frac{#1}{(z-w)^{3}}}
\newcommand{\zwtw}[1]{\frac{#1}{(z-w)^2}}
\newcommand{\zwon}[1]{\frac{#1}{z-w}}
\begin{document}
\renewcommand{\thefootnote}{\fnsymbol{footnote}}
\begin{titlepage}
\null
\begin{flushright}
NBI-HE-92-42 \\
July 1992
\end{flushright}
\vspace{3cm}
\begin{center}
{\LARGE
Free Field Representations \\
of Extended Superconformal Algebras
\par}
\vskip 3em
\lineskip .75em
\normalsize
\begin{tabular}[t]{c}
{\large Katsushi Ito} \\
{\large Jens Ole Madsen} \\
{\large Jens Lyng Petersen}  \\ \\
{\sl The Niels Bohr Institute} \\
{\sl University of Copenhagen} \\
{\sl Blegdamsvej 17, DK-2100}  \\
{\sl Copenhagen {\O}, Denmark}
\end{tabular}
\vskip 1.5em
{\bf Abstract}
\end{center} \par
We study the classical and quantum $G$ extended superconformal
algebras from the hamiltonian reduction of affine Lie superalgebras,
with even subalgebras $G\oplus sl(2)$.
At the classical level we obtain generic formulas for the Poisson
bracket structure of the algebra.
At the quantum level we get free field (Feigin-Fuchs) representations
of the algebra by using the BRST formalism and the free field
realization of the affine Lie superalgebra.
In particular we get the free field representation of the
$sl(2)\oplus sp(2N)$ extended superconformal algebra from the Lie
superalgebra $osp(4|2N)$.
We also discuss the screening operators of the algebra and  the
structure of singular vectors in the free field representation.

\end{titlepage}
\baselineskip=0.8cm
\renewcommand{\thefootnote}{\arabic{footnote}}
\setcounter{footnote}{0}

\sect{Introduction}
The interplay of the world-sheet superconformal symmetry and the
target space symmetry of related supersymmetric non-linear $\sigma$
models has attracted much attention in relation to the
compactification of superstrings and topological field theories.
In particular, the relation between $N=2$ and $4$ superconformal
algebras and the supersymmetric non-linear $\sigma$ models on
Calabi-Yau manifolds is well known and has been studied
in great detail \cite{n2},\cite{n4}.
It seems an interesting problem to study a further extension of the
superconformal algebra as well as
its geometrical structure, which might have further nontrivial
topological properties.

There are several distinct approaches to the construction of extended
superconformal algebras.
Originally, the linear $o(N)$ supersymmetric extension of the Virasoro
algebra, which was introduced by Ademollo et al. \cite{Ade}, was
realized as the superconformal transformation on extended superspace
with internal $o(N)$ symmetry.
However, this formulation is valid only for $N\leq 4$ due to the
appearance of negative conformal weight supercurrents \cite{Ade} and the
absence of a central extension for $N\geq 5$ \cite{Sch}.
A slightly different approach is to regard linear $N=2,4$ extended
superconformal algebras as a symmetry of supersymmetric non-linear
$\sigma$ models on group manifolds and on a coset space such as a hermitian
symmetric space \cite{KaSu} or the Wolf spaces \cite{sigma}.

Knizhnik and Bershadsky \cite{KnBe} have proposed a {\em non-linear}
extension of superconformal algebras with $u(N)$ and $so(N)$
Kac-Moody symmetries by considering the closure and the associativity
of operator product expansions for the currents (the OPE method).
This non-linearity means that these extended superconformal algebras
belong to the class of Zamolodchikov's $W$-algebras \cite{Zamo}.
It is by now well understood that the $W$-algebra associated with
a simple Lie algebra can be obtained from the hamiltonian reduction
\cite{DrSo}-\cite{BaFeFoRaWi} of the corresponding affine Lie algebra by
considering the constraints on the phase space of currents using a
gauge fixing procedure.
Recently various types of extensions of the $W$-algebra
\cite{Po}-\cite{Rom}
including their supersymmetric generalization \cite{Kh}-\cite{superW} have
been considered by using the hamiltonian reduction technique.
Among them, in refs. \cite{BaTjDr} and \cite{Rom}, the $u(N)$ bosonic
superconformal algebras have been constructed from a certain hamiltonian
reduction procedure of the affine $sl(N+2)$ algebra.

In ref. \cite{ItMa}, two of the present authors have proposed a procedure
for constructing the classical extended superconformal algebras by the
hamiltonian reduction of affine Lie superalgebras with even Lie subalgebras
$G\oplus sl(2)$ (see table \ref{ta:li}).
Remarkably our classification corresponds to that of the reduced holonomy
groups of non-symmetric Riemannian manifolds \cite{Ber}.
We expect that these extended superconformal algebras might correspond to
non-linear $\sigma$ models with rich geometrical structures such as
quaternionic or octonionic structures.
Recently a similar classification based on the OPE method has been
proposed in \cite{BoFrLi}.

The purpose of the present paper is to develop the hamiltonian
reduction in the quantum case and to study the representation theory
of the extended superconformal algebra using free fields, which is
useful for the analysis of singular vectors and the computation of
correlation functions \cite{FeFu}.

This paper is organized as follows:
In sect. 2, after we review some basic properties of affine Lie
superalgebras, we discuss the classical hamiltonian reduction
for an affine Lie superalgebra $\HBG$ associated with a Lie
superalgebra $\BG$ which has the even subalgebra $G\oplus sl(2)$ and we
derive the classical $G$ extended superconformal algebra.
In sect. 3, by using the BRST gauge fixing procedure and the
Wakimoto realization \cite{FeFr} of the affine Lie superalgebra, we derive
the free field realization of the extended superconformal algebra.
In sect 4, some explicit examples of extended superconformal algebras
corresponding to the non-exceptional Lie superalgebras are given.
In sect. 5, we investigate the structure of screening operators and
the null field structure of degenerate representations of
the $G$ extended superconformal algebra.

\sect{The Classical Hamiltonian Reduction}
\subsect{Affine Lie Superalgebras}
We start with explaining some definitions of basic classical Lie
superalgebras \cite{Kac} and their affine extensions.
Let $\BG=\BG_{0}\oplus\BG_{1}$ be a rank $n$ basic classical Lie
superalgebra with an even subalgebra $\BG_{0}$ and an odd subspace
$\BG_{1}$.
$\DE=\DE^{0}\cup\DE^{1}$ is the set of roots of $\BG$, where
$\DE^{0}$ ($\DE^{1}$) is the set of even (odd) roots.
Denote  the set of positive even (odd) roots by
$\DE^{0}_{+}$ ($\DE^{1}_{+}$).
The superalgebra $\BG$ has a canonical basis
$\{ E_{\A}, e_{\G}, h^{i} \}$ ($\A\in\DE^{0}$, $\G\in\DE^{1}$,
$i=1, \ldots ,n$), which satisfies (anti-)commutation relations
\eqn
\mbox{[} E_{\A}, E_{\B} \mbox{]}&=&
\left\{ \begin{array}{ll}
N_{\A,\B} E_{\A+\B}, &\quad \mbox{ for $\A$, $\B$, $\A+\B\in \DE^{0}$},\\
\displaystyle{{2\A\cdot h \over \A^{2}}}, &\quad \mbox{ for $\A+\B=0$, }
\end{array}\right.  \CR
\left\{ e_{\G}, e_{\G'} \right\}&=& N_{\G,\G'} E_{\G+\G'}, \quad
\mbox{ for $\G, \G'\in \DE^{1}$, $\G+\G'\in \DE^{0}$},  \CR
\left\{ e_{\G}, e_{-\G} \right\}&=& \G\cdot h ,
\quad \mbox{for $\G\in\DE^{1}_{+}$}, \CR
\mbox{[} e_{\G}, E_{\A} \mbox{]}&=& N_{\G,\A} e_{\G+\A}, \quad
\mbox{ for $\A\in \DE^{0}$, $\G$,$\G+\A\in \DE^{1}$},  \CR
\mbox{[} h^{i}, E_{\A} \mbox{]}&=&\A^{i}E_{\A}, \quad
\mbox{[} h^{i}, e_{\G} \mbox{]}=\G^{i}e_{\G}.
\enn
The even subalgebra $\BG_{0}$ is generated by $\{ E_{\A}, h^{i} \}$.
The odd subspace $\BG_{1}$ is spanned by $e_{\G}$.
$\BG_{0}$ acts on $\BG_{1}$ as a faithful representation.
The Killing form $( \ ,\ )$ on $\BG$ is defined by
\eq
(E_{\A}, E_{\B})={2\over \A^{2}}\D_{\A+\B,0}, \quad
(e_{\G}, e_{-\G'})=-(e_{-\G}, e_{\G'})=\D_{\G,\G'},  \quad
(h^{i},h^{j})=\D_{i j},
\en
for $\A,\B\in\DE^{0}$, $\G,\G' \in \DE^{1}_{+}$, $i,j=1,\ldots, n$.

An affine Lie superalgebra $\HBG$ at level $k$ is the
(untwisted) central extension of $\BG$ and consists of the elements
of the form $(X(z), x_{0})$, where $X(z)$ is a $\BG$-valued
Laurent polynomial of $z\in \BC$ and $x_{0}$ is a number
\cite{reimansemenov}.
The commutation relation for two elements $(X(z), x_{0})$ and
$(Y(z), y_{0})$ is given by
\eq
\mbox{[} \ (X(z), x_{0}), \ (Y(z), y_{0}) \ \mbox{]}
=(\ \mbox{[}\ X(z), \ Y(z) \ \mbox{]},\ k \oint \dz (X(z), \pa Y(z))\ ).
\label{eq:com}
\en
The dual space $\HBG^{*}$ of $\HBG$ is generated by the current
$(J(z), a_{0})$.
Using the Killing form $( \ , \ )$ on $\BG$, we may identify
$\HBG^{*}$ with $\HBG$.
The inner product $\langle\ , \ \rangle$ of $(J(z), a_{0}) \in\HBG^{*}$
and $(X(z), x_{0}) \in\HBG$ is given by
\eq
\langle \ (J , a_{0}), \ (X, x_{0}) \ \rangle
=\oint\dz (\ J(z),\  X(z)\ ) +a_{0}x_{0}.
\en
Let us define the coadjoint action $\ad^{*}$ of $\HBG$
on $\HBG^{*}$ by
\eq
\langle\ \ad^{*}(X, x_{0})\ (J, a_{0}),\ (Y, y_{0})\ \rangle
=-\langle\ (J, a_{0}),\
\mbox{[}\ (X, x_{0}),\ (Y, y_{0})\ \mbox{]}\ \rangle.
\en
Using (\ref{eq:com}), we get
\eq
\ad^{*}(X, x_{0})(J(z), a_{0})
=(\ \mbox{[}\ X(z), J(z)\ \mbox{]}+k a_{0}\pa X(z),\ 0\ ).
\en
Namely, the coadjoint action is nothing but the infinitesimal gauge
transformation of the current $J(z)$.
Denote this gauge transformation with the gauge parameter $\LM(z)$
as $\D_{\LM}$:
\eq
\D_{\LM} J(z)=\mbox{[} \ \LM(z),\  J(z)\ \mbox{]} +k \pa \LM (z),
\label{eq:gauge1}
\en
In the following we take the number $a_{0}$ to be 1.
In terms of the canonical basis
\eq
J(z)=\sum_{\A\in\DE^{0}}{\A^{2}\over 2}J_{\A}(z)E_{\A}
     +\sum_{\G\in\DE^{1}}j_{\G}(z)e_{\G}+\sum_{i=1}^{n}H^{i}(z)h^{i}.
\label{eq:current}
\en
and
\eq
\LM(z)=\sum_{\A\in\DE^{0}}\vep_{\A}(z)E_{\A}
     +\sum_{\G\in\DE^{1}}\xi_{\G}(z)e_{\G}
     +\sum_{i=1}^{n}\vep^{i}(z)h^{i},
\en
we can express the gauge transformation (\ref{eq:gauge1}) in terms
of components:
\eqn
\D_{\LM} J_{\A}\!\!\!\! &=& \!\!\!\! \sum_{\B, \A-\B\in \DE^{0}}
             {(\A-\B)^{2}\over \A^{2}}N_{\B,\A-\B} \vep_{\B}J_{\A-\B}
            -{2 (\A\cdot H) \over \A^{2}} \vep_{\A}
            +{2k\over \A^{2}}\pa\vep_{\A}
            +(\A\cdot\vep) J_{\A} \CR
         & & +\sum_{\G,\A-\G\in \DE^{1}}
             {2\over \A^{2}}N_{\G,\A-\G} \xi_{\G}j_{\A-\G}, \CR
\D_{\LM} j_{\G} \!\!\!\! &=& \!\!\!\!
             \sum_{\AT{\A\in\DE^{0}}{\G-\A\in\DE^{1}}}
             \!\!\!\! N_{\A,\G-\A}\vep_{\A}j_{\G-\A}
            +(\G\cdot\vep)j_{\G}
             +\!\!\!\! \sum_{\AT{\A\in\DE^{0}}{\G-\A\in\DE^{1}}}
       \!\!\!\! {\A^{2}\over 2}N_{\G-\A,\A}\xi_{\G-\A}J_{\A}
             -\xi_{\G} \G\cdot H +k\pa\xi_{\G}, \CR
\D_{\LM} H^{i} \!\!\!\! &=& \!\!\!\!
             \sum_{\A\in\DE^{0}}\A^{i}\vep_{\A}J_{-\A}
             +\sum_{\G\in\DE^{1}_{+}}
              \G^{i}(\xi_{\G}j_{-\G}+\xi_{-\G}j_{\G})
             +k\pa\vep^{i}.
\label{eq:gauge2}
\enn
Writing the gauge transformations $\D_{\LM}$ as
\eqn
\D_{\LM}&=&\oint\dz (\LM(z), J(z)) \CR
        &=&\oint\dz \sum_{\A\in\DE^{0}}\vep_{\A}J_{-\A}
           +\sum_{\G\in\DE^{1}_{+}}(j_{\G}\xi_{-\G}+\xi_{\G}j_{-\G})
           +\sum_{i=1}^{n}\vep^{i}H^{i},
\enn
one can introduce a canonical Poisson structure on the dual space
$\HBG^{*}$.
This Poisson structure is expressed in the form of the operator product
expansions:
\eqn
J_{\A}(z)J_{\B}(w)&=&\left\{
\begin{array}{ll}
\displaystyle{\ordo{N_{\A,\B}J_{\A+\B}(w)}+\cdots} ,& \quad
\mbox{ for $\A$, $\B$, $\A+\B\in \DE^{0}$}, \\
 \displaystyle{\ord{2}{{2k\over \A^{2}}}
 +\ordo{ {2\A\cdot H(w)\over \A^{2}} }+\cdots}, &\quad
\mbox{ for $\A+\B=0$,}
\end{array}\right. \CR
j_{\pm\G}(z)j_{\pm\G'}(w)&=&
\ordo{N_{\mp\G,\mp\G'}J_{\pm(\G+\G')}(w)}+\cdots ,\quad
\mbox{ for $\G, \G'\in \DE^{1}_{+}$, $\G+\G'\in \DE^{0}$},  \CR
j_{\G}(z)j_{-\G'}(w)&=&\left\{
\begin{array}{ll}
\displaystyle{\ordo{-N_{\G',-\G}J_{\G-\G'}(w)}+\cdots} ,& \quad
\mbox{ for $\G,\G'\in\DE^{1}_{+}$, $\G-\G'\in\DE^{0}$}, \\
\ord{2}{-k}+\ordo{-\G\cdot H(w)}+\cdots,
&\quad \mbox{ for $\G=\G'\in\DE^{1}_{+}$,}
\end{array}\right.  \CR
J_{\A}(z)j_{\G}(w)&=&{N_{-\A,\G+\A}j_{\G+\A}(w)\over z-w}+\cdots ,\quad
\mbox{ for $\G\in \DE^{1}$, $\A\in \DE^{0}$, },  \CR
H^{i}(z)J_{\A}(w)&=&{\A^{i}J_{\A}(w)\over z-w}+\cdots , \quad
H^{i}(z)j_{\G}(w)={\G^{i}j_{\G}(w)\over z-w}+\cdots. \CR
H^{i}(z)H^{j}(w)&=&{k\D^{i j}\over (z-w)^{2}}+\cdots .
\enn
Here we use some identities for the structure constants, which come
from the Jacobi identities:
\eqn
{2N_{\A,\B}\over (\A+\B)^{2}}&=&{2 N_{\B,-\A-\B}\over \A^{2}}
                               ={2 N_{-\A-\B,\A}\over \B^{2}},
\quad \mbox{for $\A,\B \in\DE^{0}$, }\CR
{2N_{\G,\G'}\over (\G+\G')^{2}}&=&
N_{\G',-\G-\G'}=-N_{-\G-\G',\G}, \CR
{2N_{\G,-\G'}\over (\G-\G')^{2}}&=&
N_{-\G',-\G+\G'}=N_{-\G+\G',\G}, \quad
\mbox{for $\G,\G'\in\DE^{1}_{+}$ }.
\label{eq:str}
\enn
In the following we will study a class of Lie superalgebras,
which include even algebras of the form $G\oplus A_{1}$.
Using Kac's notation, these algebras are classified as follows
$A(n|1)\ (n\geq 1)$, $A(1,0)=C(2)$, $B(n|1)\ (n\geq 0)$,
$B(1|n)\ (n\geq 1)$, $D(n|1)\ (n\geq 2)$, $D(2|n)\ (n\geq 1)$,
$D(2|1;\A)$, $F(4)$ and $G(3)$ (see table \ref{ta:li}).
The embedding of $A_{1}$ in $\BG$ carried by ${\bf g_1}$
has spin $\half$, except for $B(1|n)$.
In the case of $B(1|n)$ the embedding has spin one.
In the spin $\half$ embedding case,
the odd subspace $\BG_{1}$ belongs to the spin $\half$ representation
with respect to the even subalgebra $A_{1}$:
\eq
\BG_{1}=(\BG_{1})_{+\half}\oplus(\BG_{1})_{-\half}.
\label{eq:dec}
\en
Based on the decomposition (\ref{eq:dec}), the odd root space $\DE^{1}$
can be divided into two parts
\eq
\DE^{1}=\DE^{1}_{+\half}\cup\DE^{1}_{-\half},
\en
where $(\BG_{1})_{\pm\half}$ is spanned by the generators whose roots
belong to $\DE^{1}_{\pm \half}$.
More explicitly, the sets $\DE^{1}_{\pm \half}$ consist of the roots
$\G\in \DE^{1}$ satisfying
\eq
{\G \cdot \theta \over \theta^{2}}=\pm\half,
\label{eq:odd1}
\en
where $\theta$ is the simple root of $A_{1}$.
By choosing appropriate simple roots for $\BG$, we can take the root
space $\DE^{1}_{+\half}$  as the space of odd positive roots;
\eq
\DE^{1}_{+}=\DE^{1}_{+\half}.
\en
Moreover we find that the relation
\eq
\DE^{1}_{+\half}=\DE^{1}_{-\half}+\theta,
\label{eq:odd2}
\en
holds.
We note that each odd space $(\BG_{1})_{\pm\half}$ belongs to a
fundamental representation of $G$ of dimension $|\DE^{1}_{+}|$, this
being the number of positive, odd roots.
{}From the explicit expressions for the root system which is given in
appendix A, we will find that the root system of the even subalgebra of
$G$ is expressed in the form of $\G-\G'$, where $\G$ and $\G'$ are
some positive odd roots.
This fact turns out to be essential for the study of
the structure of general $G$
extended superconformal algebras, as will be discussed in the
next subsection.

\subsect{Classical hamiltonian reduction}
We consider the hamiltonian reduction of the phase space of the
currents $\HBG^{*}$ associated with the Lie superalgebra $\BG$
with the even subalgebra $G\oplus A_{1}$.
Our main idea is to impose the constraint on the even subalgebra
$A_{1}$ and to keep affine $G$ algebra symmetry.
For $A(1,0)$, this reduction procedure has been discussed in
ref. \cite{Kh}.
We start from  the phase space $\cF$ of the currents with the constraint
\eq
J_{-\theta}(z)=1 .
\label{eq:cons1}
\en
Let us consider the gauge group $\cN$ which preserves the constraint
(\ref{eq:cons1}).
Putting $-\theta$ for $\A$ in the first formula of eqs. (\ref{eq:gauge2}),
we get
\eq
\D_{\LM} J_{-\theta}=
            {2 (\theta\cdot H) \over \theta^{2}} \vep_{-\theta}
            +{2k\over \theta^{2}}\pa\vep_{-\theta}
            -(\theta\cdot\vep) J_{-\theta}
         +\sum_{\G \in \DE^{1}_{+}}
     {2\over \theta^{2}}N_{-\G,-\theta+\G} \xi_{-\G}j_{-\theta+\G}.
\label{eq:gau1}
\en
Here we use the relation (\ref{eq:odd2}) and $N_{\A, -\theta-\A}=0$
for  $\A\in\DE(G)$.
Therefore the condition which preserves the constraints
(\ref{eq:cons1}) $\D_{\LM}J_{-\theta}=0$ is equivalent to
\eq
\vep_{-\theta}=\theta\cdot\vep=0, \quad
\xi_{-\G}=0, \quad {\rm for } \ \G\in\DE^{1}_{+}.
\en
Namely the Lie algebra $\Bn$ of the gauge group $\cN$ is
equal to $\WG\oplus \Bn_{1}$, where $\WG$ is
the affine Lie algebra corresponding to the even subalgebra $G$ and
$\Bn_{1}$ is generated by the elements:
\eq
\LM(z)=\vep_{\theta}(z)E_{\theta}
     +\sum_{\G\in\DE^{1}_{+}}\xi_{\G}(z)e_{\G}.
\en
Let $\cN_{1}$ be the gauge group corresponding to $\Bn_{1}$.
The reduced phase space $\cM$ is defined as the quotient space
$\cM=\cF/\cN_{1}$.
By the standard gauge fixing procedure, one chooses one or the other gauge
slice as a representative of the reduced phase space.
One typical gauge is the \lq\lq Drinfeld-Sokolov (DS)"-type gauge
\cite{DrSo}:
\eq
J_{\theta}(z)=T (z), \quad
\theta \cdot H(z)=0, \quad
j_{\G}(z)=G_{\G}(z), \quad
j_{-\G}(z)=0,
\label{eq:dsgauge}
\en
for $\G\in\DE^{1}_{+}$.
The generic gauge transformation projected on the DS gauge slice
becomes in general the transformation corresponding to the extended
Virasoro algebra \cite{DrSo}-\cite{BeOo2}.
Hence the final step is to consider the generic gauge transformation
preserving the gauge condition (\ref{eq:dsgauge}). In particular
\eqn
\D_{\LM}J_{-\T}\!\!\!\! &=&\!\!\!\! -(\T\cdot\vep)
      +{2k\over \T^{2}}\pa\vep_{-\T}, \CR
\D_{\LM}(\T\cdot H)\!\!\!\! &=& \!\!\!\!
      \T^{2}\vep_{\T}-\T^{2}\vep_{-\T} T
      +k\pa(\T\cdot\vep)
      +\sum_{\G\in\DE^{1}_{+}}\T\cdot\G\xi_{-\G}G_{\G}, \\
\D_{\LM}j_{-\G}\!\!\!\! &=& \!\!\!\!
      N_{-\T,\T-\G}\vep_{-\T}G_{\T-\G}
      +\!\!\!\! \sum_{\AT{\G'\in\DE^{1}_{+}}{\G'-\G\in\DE(G)}}
       \!\!\!\! N_{\G,-\G'}\xi_{-\G'}J_{\G'-\G}
      +N_{\T-\G,\G}\xi_{\T-\G}
      +\xi_{-\G}(\G\cdot H)+k\pa\xi_{-\G},  \nonumber
\enn
for $\G\in\DE^{1}_{+}$.
By solving the conditions which preserve the DS-gauge
$\D J_{-\T}=0$, $\D (\T\cdot H)=0$ and $\D j_{-\G}=0$,
we can express the parameters $\vep_{\theta}$, $\theta\cdot\vep$ and
$\xi_{\G}$ ($\G\in\DE^{1}_{+}$) in terms of the
other parameters, $\vep_{-\theta}$, $\xi_{-\G}$ ($\G\in\DE^{1}_{+}$),
$\vep_{\A}$ and $\A\cdot \vep$ ($\A \in \DE(G)$):
\eqn
\T\cdot\vep\!\!\!\!&=& \!\!\!\! {2k\over \T^{2}}\pa\vep_{-\T},
\quad \quad \quad
\vep_{\T}=\vep_{-\T} T
          -{2k^{2}\over (\T^{2})^{2}}\pa^{2}\vep_{-\T}
          -\half \sum_{\G\in\DE^{1}_{+}}\xi_{-\G}G_{\G},
\label{eq:gaugetr1} \\
\xi_{\T-\G}\!\!\!\!&=& \!\!\!\!
           {-1\over N_{\T-\G,\G}}\bigl\{
           N_{-\T,\T-\G}G_{\T-\G}\vep_{-\T}
           +\!\!\!\! \sum_{\AT{\G'\in\DE^{1}_{+}}{\G-\G'\in\DE(G)}}
            \!\!\!\!N_{\G,-\G'}\xi_{-\G'}J_{\G'-\G}
           +\xi_{-\G}(\G\cdot H)+k\pa\xi_{-\G} \bigr\}, \nonumber
\enn
In the DS gauge,
the gauge transformation of $J_{\theta}(z)=T(z)$
becomes
\eq
\D T=\sum_{\G\in\DE^{+}}{2\over\T^{2}}N_{\G,\T-\G}\xi_{\G}G_{\T-\G}
     +(\T\cdot\vep)T+{2k\over \T^{2}}\pa\vep_{\T}.
\en
{}From (\ref{eq:gaugetr1}) we get
\eqn
\D T&=& -{4k^{3}\over (\T^{2})^{3}}\pa^{3}\vep_{-\T}
      +{2k\over\T^{2}}(2T\pa\vep_{-\T}+\vep_{-\T}\pa T)
\label{eq:emte} \\
   & & -{1\over\T^{2}}\sum_{\G\in\DE^{1}_{+}}
\bigl\{  3k \pa\xi_{-\G}G_{\G}+k\xi_{-\G}\pa G_{\G}
        +2\xi_{-\G}(\G\cdot H)G_{\G}
      +2\!\!\!\! \sum_{\AT{\G'\in\DE^{1}_{+}}{\G'-\G\in\DE(G)}}
\!\!\!\!\!\!  N_{\G,-\G'}\xi_{-\G'}G_{\G}J_{\G'-\G}
\bigr\} .  \nonumber
\enn
With respect to the parameter $\vep_{-\theta}$, $T(z)$ behaves
as an energy momentum tensor under conformal transformations.
The $G$ Kac-Moody currents  $J_{\B}$ and $\B\cdot H$ ($\B\in\DE(G)$)
transform as
\eqn
\!\!\!\!\!\!\!\!
\D J_{\B}\!\!\!\! &=& \!\!\!\! \!\!\!\!
\sum_{\A,\B-\A\in\DE(G)} N_{-\A,\B}\vep_{\A}J_{\B-\A}
           -{2\B\cdot H \over \B^{2}}\vep_{\B}
           +{2k\over\B^{2}}\pa\vep_{\B}
           +(\B\cdot\vep)J_{\B}
           +\!\!\!\!\sum_{\G,\B-\G\in \DE^{1}_{+}}\!\!\!\!
                {2N_{-\G,\B+\G}\over\B^{2}}\xi_{-\G}G_{\B+\G}, \CR
\D (\B\cdot H)\!\!\!\! &=&\!\!\!\!
\sum_{\A\in\DE(G)}(\B\cdot\A)\vep_{\A}J_{-\A}
                +k\pa(\B\cdot\vep)
                -\sum_{\G\in\DE^{1}_{+}}(\B\cdot\G)\xi_{-\G}G_{\G}.
\label{eq:curr}
\enn
The gauge transformation of supercurrents $G_{\G}(z)$
($\G\in \DE^{1}_{+}$) in the DS gauge is given by
\eqn
\D G_{\G}(z)&=&\sum_{\AT{\A\in\DE(G)}{\G-\A\in\DE^{1}_{+}}}
                N_{\A,\G-\A}\vep_{\A}G_{\G-\A}
              +(\G\cdot\vep)G_{\G} \label{eq:super}\\
            & & -\sum_{\AT{\G'\in\DE^{1}_{+}}{\G'-\G\in\DE(G)}}
                N_{\G',-\G}\xi_{\G'}J_{-\G'+\G}
              -N_{-\G,\G-\T}\xi_{\G-\T}T
              -\xi_{\G}(\G\cdot H)+k\pa\xi_{\G}.  \nonumber
\enn
By inserting (\ref{eq:gaugetr1}) into this expression we get the
transformation on the reduced phase space.
The parts in (\ref{eq:super}) containing $\vep_{-\theta}$,
$\vep^{i}$ and  $\vep_{\A}$ ($\A\in\DE(G)$) are consistent with
(\ref{eq:emte}) and (\ref{eq:curr}).
The nontrivial gauge transformation is that containing
the anticommuting gauge parameters $\xi_{-\G'}$, which is denoted by
$\D_{\xi}G_{\G}$.
{}From (\ref{eq:super}) and (\ref{eq:gaugetr1}), one finds that this
transformation becomes
\eqn
\D_{\xi}G_{\G}
&=&\!\!\!\! {-k^{2}\over N_{\G,\T-\G}}\pa^{2}\xi_{\G-\T}
+\sum_{ \AT{\G'\in\DE^{1}_{+}}{\G'-\G\in\DE(G)}}
 2{N_{\G',-\G}\over N_{\G',\T-\G'}}k\pa\xi_{\G'-\T}J_{\G-\G'}
+{2\over N_{\G,\T-\G}}k\pa\xi_{\G-\T}(\G\cdot H) \CR
& & \!\!\!\! -N_{-\G,\G-\T}\xi_{\G-\T}T
+{k\xi_{\G-\T} \pa(\G\cdot H)  \over N_{\G, \T-\G}}
-{\xi_{\G-\T}(\G\cdot H)^{2}\over N_{\G, \T-\G}}   \CR
& &\!\!\!\! + \sum_{\AT{\G'\in\DE^{1}_{+}}{-\G'+\G\in\DE(G)}}
  {N_{\G',-\G}\over N_{\G',\T-\G'}}\xi_{-\T+\G'}k\pa J_{\G-\G'}
 +\sum_{\AT{\G'\in\DE^{1}_{+}}{\G'-\G\in\DE(G)}}
  {N_{\G',-\G}\over N_{\G',\T-\G'}}\xi_{\G'-\T}J_{-\G'+\G}
  (\G'-\G)\cdot H \CR
& & \!\!\!\! +\sum_{\AT{\G'\in\DE^{1}_{+}}{\G'-\G\in\DE(G)}}
   \sum_{\AT{\G''\in\DE^{1}_{+}}{-\G'+\G''\in\DE(G)}}
   {N_{\G',-\G}N_{\T-\G',\G''-\T}\over N_{\G',\T-\G'}}
    \xi_{-\T+\G''}J_{\G-\G'}J_{\G'-\G''},
\enn
Here we have used the identities for the structure constants
eqs.(\ref{eq:str}), as well as
\eq
{N_{\T-\G,\G'-\T}\over N_{\G,\T-\G}}=-{N_{\G',-\G}\over N_{\G',\T-\G'}},
\en
for $\G,\G'\in\DE^{1}_{+}$ in order to simplify the formula.
This formula can be checked by using explicit matrix representations
of $\BG$ \cite{Cor},\cite{DeNi} in table \ref{ta:li}.
If we denote the gauge transformation in the DS-gauge as
\eq
\D=\oint\dz (\LM(z)\ ,J_{DS}(z)),
\en
where
\eq
J_{DS}(z)= T(z)E_{\theta}
          +\sum_{\G\in \DE^{1}_{+}}G_{\G}(z)e_{\G}
          +\sum_{\A\in\DE(G)} J_{\A}(z) E_{\A}
          +\sum_{i=1}^{\rank (G)}H^{i}h^{i},
\en
we get the Poisson bracket structure on the reduced phase space.
Here we take $h^{i}$ ($i=1, \ldots, \rank (G)$) as the generators of the
Cartan subalgebra of $G$.
After rescaling $\TT={\T^{2}\over 2k}T$ and
$\TG_{\G}=\sqrt{{N_{\T-\G,\G}\over k}}G_{\G}$ for
$\G\in\DE^{1}_{+}$,
these classical Poisson brackets may be written as follows in the form of
formal operator product expansions:
\eqn
\TT(z)\TT(w)&=&\ord{4}{{-6k\over \T^{2}}}
         +\ord{2}{2\TT(w)}
         +\ordo{\pa \TT(w)}+\cdots, \CR
\TT(z)\TG_{\G}(w)&=&\ord{2}{{3\over2}\TG_{\G}(w)}+\ordo{\pa\TG_{\G}(w)} \CR
    & &  -\ordo{ {1\over k}\bigl\{  (\G\cdot H)\TG_{\G}
+\sum_{\AT{\G'\in\DE^{1}_{+}}{\G-\G'\in\DE(G)}}
                 N_{\G,-\G'}\TG_{\G'}J_{\G-\G'}(w) \bigr\} }+\cdots, \CR
J_{\B}(z)\TG_{\G}(w)&=&\ordo{-N_{-\B,-\G}\TG_{\B+\G}(w)}+\cdots, \quad
H^{i}(z)\TG_{\G}(w)=\ordo{\G^{i}\TG_{\G}(w)}+\cdots, \CR
\TG_{\T-\G'}(z)\TG_{\G}(w)&=&\ord{3}{2 k\D_{\G',\G}}
+\ord{2}{{\cal O}^{2}_{\G',\G}(w)}+\ordo{{\cal O}^{1}_{\G',\G}(w)}
+\cdots,
\label{eq:clope}
\enn
where
\eqn
{\cal O}^{2}_{\G',\G}&=& \!\!\!\! \left\{
\begin{array}{ll}
-2N_{\G',-\G} \sqrt{{N_{\G,\T-\G} \over N_{\G',\T-\G'}}}J_{\G-\G'}, &
\mbox{for $\G-\G'\in\DE(G)$} \\
-2\G\cdot H, & \mbox{for $\G'=\G$},
\end{array}
\right. \\
{\cal O}^{1}_{\G',\G}&=& \!\!\!\! \left\{
\begin{array}{ll}
-N_{\G',-\G} \sqrt{{N_{\G,\T-\G} \over N_{\G',\T-\G'}}}\pa J_{\G-\G'}
+{1\over k} \sqrt{{N_{\G',\T-\G'} \over N_{\G,\T-\G}}}
N_{\G',-\G'} J_{\G-\G'}(\G'-\G)\cdot H  &  \\
\quad  +{1\over k} \sqrt{{N_{\G,\T-\G} \over N_{\G',\T-\G'}}}
\!\!\!\!
\displaystyle{\sum_{\AT{\G''\in\DE^{1}_{+}}{\G''-\G',\G''-\G\in\DE(G)}}
   \!\!\!\! N_{\G'',-\G}N_{\G'',-\G'}J_{-\G'+\G''}J_{-\G''+\G}},
& \!\!\!\! \!\!\!\! \mbox{for $\G-\G'\in \DE(G)$,} \\
{2\over \T^{2}}N_{-\T+\G,-\G}N_{\T-\G,\G}\TT
-\pa(\G\cdot H)+{1\over k}(\G\cdot H)^{2} & \\
\quad +{1\over k} \!\!\!\!
 \displaystyle{\sum_{\AT{\G''\in\DE^{1}_{+}}{\G''-\G\in\DE(G)}}
   \!\!\!\! (N_{\G'',-\G})^{2}J_{\G-\G''}}J_{\G''-\G},
& \!\!\!\! \!\!\!\!
\mbox{for $\G'=\G$}.
\end{array} \right.
\enn
Note that the supercurrents $\TG_{\G}$ are not primary fields with respect
to $\TT(z)$.
However, if we define the total energy-momentum tensor
$T_{ESA}$ by adding the Sugawara form $T_{Sugawara}$ of the
$\WG$ affine Lie algebra: $T_{ESA}=\TT+T_{Sugawara}$ where
\eq
T_{Sugawara}={1\over 2k}\left\{
\sum_{\A\in\DE(G)} {\A^{2}\over 2}J_{\A}J_{-\A}
     +\sum_{i=1}^{\rank (G)}H^{i}H^{i} \right\},
\en
we can check that the supercurrents have conformal weight 3/2 with
respect to $T_{ESA}$.
The classical value of the central charge $c_{ESA}$ is
$-12k/\T^{2}$.

\sect{The Quantum Hamiltonian Reduction}
\subsect{BRST formalism}
In this section we discuss the quantum hamiltonian reduction \cite{BeOo}
for an affine Lie superalgebra $\HBG$ at level $k$, generated by
$J_{\A}(z)$ ($\A\in\DE^{0}$), $j_{\G}(z)$ ($\G\in\DE^{1}$)
and $H^{i}(z)$ ($i=1, \ldots, n$).
Let $T_{WZNW}(z)$ be the energy-momentum tensor of an affine Lie
superalgebra $\HBG$, defined by the Sugawara form:
\eq
\!\!\!\!\! T_{WZNW}={1\over 2(k+h^{\vee})}
\left( :\sum_{\A\in\DE^{0}_{+}}{\A^{2}\over2}(J_{\A}J_{-\A}+J_{-\A}J_{\A})
       +\sum_{\G\in\DE^{1}_{+}}(j_{\G}j_{-\G}-j_{-\G}j_{\G})
       +\sum_{i=1}^{n}H^{i}H^{i} : \right),
\label{eq:sug}
\en
where $h^{\vee}$ is the dual Coxeter number of $\BG$ and  $: \ :$ denotes
the normal ordering.
In order to impose the constraint for the currents at the quantum
level, we have to ``improve" the energy-momentum tensor by a contribution
from the Cartan currents $H^{i}(z)$:
\eq
T_{improved}(z)=T_{WZNW}(z)-\mu\cdot\pa H(z).
\label{eq:imp}
\en
Here $\mu$ is an $n$-vector.
With respect to the improved energy-momentum tensor $T_{improved}(z)$,
the currents corresponding to the roots $\A$ have conformal
weights $1+\mu\cdot\A$.
We are concerned with a class of Lie superalgebras whose even subalgebras
are $G \oplus A_{1}$, where $G$ is a semisimple Lie algebra.
In the previous section we have considered the constraint $J_{-\T}(z) = 1$
but the $G$ currents have conformal dimension one, so in order for the
constraint to make sense, it must have improved conformal dimension, 0.
This means that  the vector $\mu$ should satisfy
\eq
\mu\cdot \T = 1, \quad
\mu\cdot \A = 0, \quad \mbox{for $\A \in \DE (G) $}.
\label{eq:mu}
\en
These equations (\ref{eq:mu}) determine the vector $\mu$
uniquely:
\eq
\mu={ \T \over \T^{2}}.
\en
{}From (\ref{eq:odd1}), we find that
the fermionic currents $j_{\G}(z)$ ($j_{-\G}(z)$) for the positive
roots $\G\in\DE^{1}_{+}$ have conformal weight 3/2 (1/2).
The current $J_{\T}(z)$ has conformal weight 2.

Now we use the BRST-gauge fixing procedure.
In the previous section we took the Drinfeld-Sokolov gauge
(\ref{eq:dsgauge}) and derived the Poisson bracket structure
for the currents.
In order to study the representation of the algebra, it is
very useful to consider the free field representation.
This is realized by taking the \lq\lq diagonal" gauge
\eqn
J_{\T}(z)&=&0 , \quad
\T\cdot H(z)= a_{0} \pa\phi(z),    \CR
j_{-\G}(z)&=& \sqrt{N_{-\G,-\T+\G}}\chi_{\G}(z),
\quad \mbox{for $\G\in\DE^{1}_{+}$}.
\enn
in (\ref{eq:current}).
Here $\phi(z)$ is a free boson coupled to the
world sheet curvature and $a_{0}$ is a constant.
$\chi_{\G}(z)$ are free fermions satisfying operator
product expansions
\eq
\chi_{\G}(z)\chi_{\T-\G'}(w)={\D_{\G,\G'}\over z-w}+\cdots,
\quad \mbox{for $\G,\G' \in \DE^{1}_{+}$.}
\en
This is consistent with the OPE's of ${\bf \hat{g}}$ together with the
constraint.
Let us introduce fermionic ghosts ($b_{\T}(z)$,$c_{\T}(z)$) with
conformal weights (0,1) and bosonic ghosts
($\tilde{b}_{\G}(z)$, $\tilde{c}_{\G}(z)$) of weight
($\half$, $\half$) for $\G\in\DE^{1}_{+}$.
The BRST current $J_{BRST}(z)$  is defined as
(cf. ref. \cite{BeOo2})
\eq
J_{BRST}(z)=c_{\T} (J_{-\T}-1)+\sum_{\G\in\DE^{1}_{+}}
\tilde{c}_{\G} (j_{-\G}-\sqrt{N_{-\G,-\T+\G}}\chi_{\G})
+\half \sum_{\G\in\DE^{1}_{+}}
N_{\G,\T-\G}:\tilde{c}_{\G}\tilde{c}_{\T-\G}b_{\T}: .
\en
We can easily  show that the BRST charge $Q_{BRST}=\oint\dz J_{BRST}(z)$
satisfies the nilpotency condition $Q_{BRST}^{2}=0$.
The total energy-momentum tensor is expressed as
\eq
T_{total}(z)=T_{improved}(z)+T_{\chi}+T_{ghost}(z),
\en
where
\eqn
T_{ghost}(z)&=& :(\pa b_{\T})c_{\T}
                +\half \sum_{\G\in\DE^{1}_{+}}
               (\Tb_{\G} \pa \Tc_{\G} - (\pa \Tb_{\G}) \Tc_{\G}):, \CR
T_{\chi}(z)&=& -\half  \sum_{\G\in\DE^{1}_{+}}
      : \chi_{\T-\G}\pa \chi_{\G}:.
\enn
The central charge $c$ of the total system is computed to be
\eq
c_{total}=c_{WZNW}-12k\mu^{2}
 +\half |\DE^{1}_{+}|
 -2 -|\DE^{1}_{+}|,
\label{eq:cen}
\en
(where $|\DE^{1}_{+}|$ is still the number of positive odd roots).
The last two terms are contributions from ghost fields.
The central charge $c_{WZNW}$ of the WZNW models on the Lie
supergroup at level $k$ is given by the formula:
\eq
c_{WZNW}={k \ \sdim \BG \over k+h^{\vee}},
\en
where the super dimension $\sdim \BG$ of a Lie superalgebra $\BG$ is
defined as ${\rm dim} \BG_{0}-{\rm dim} \BG_{1}$.
The list for the corresponding central charges
is shown in table \ref{ta:ce}.
The results here are in complete agreement with
previous calculations using a variety of different techniques.
Thus, the results for $A(n|1), B(n|1), D(n|1)$ were obtained in ref.
\cite{KnBe}; the one for $D(2|1;\alpha )$ agrees with ref. \cite{sigma,dob}
(see also, \cite{ItMaPe}), the result for $D(2|n)$ and $G(3)$ agrees with
ref. \cite{BoFrLi}, and the result for $F(4)$ with that of \cite{FrLi}.
Our treatment here, however, provides a unifying framework.

In the classical limit $k\rightarrow \infty$, the expression
(\ref{eq:ene}) becomes $-12k/\T^{2}$, which agrees with the
result (\ref{eq:clope}) obtained in the previous section.

\subsect{The Free Field Representation}
We consider the free field representations of $G$ extended superconformal
algebra based on the Wakimoto construction \cite{FeFr} of the affine Lie
superalgebra $\HBG$ \cite{Ito}.
Let us introduce bosonic ghosts $(\B_{\A}(z), \G_{\A}(z))$  for
even positive roots $\A\in\DE^{0}_{+}$ with conformal dimensions (1,0),
fermionic ghosts $(\eta_{\G}(z), \xi_{\G}(z))$ for odd positive
roots $\G\in\DE^{1}_{+}$ with (unimproved)
conformal dimensions (1,0) and $n$ free
bosons $\vp(z)=(\vp_{1}(z),\ldots,\vp_{n}(z))$ coupled to the world-sheet
curvature, satisfying the operator product expansions:
\eqn
\B_{\A}(z)\G_{\A'}(w)&=&{\D_{\A,\A'}\over z-w}+\cdots, \quad
\eta_{\G}(z)\xi_{\G'}(w)={\D_{\G,\G'}\over z-w}+\cdots, \CR
\vp_{i}(z)\vp_{j}(w)&=&-\D_{i j}{\rm ln}(z-w)+\cdots.
\enn
Using these free fields the energy-momentum tensor is expressed as
\eq
T_{WZNW}(z)=\sum_{\A\in\DE^{0}_{+}}:\B_{\A}\pa\G_{\A}:
           -\sum_{\G\in\DE^{1}_{+}}:\eta_{\G}\pa\xi_{\G}:
           -\half :(\pa\vp)^{2}:-{\I\rho\cdot\pa^{2}\vp \over \A_{+}},
\en
where $\A_{+}=\sqrt{k+h^{\vee}}$ and $\rho=\rho_{0}-\rho_{1}$.
$\rho_{0}$ ($\rho_{1}$) is half the sum of positive even (odd) roots:
\eq
\rho_{0}=\half\sum_{\A\in\DE^{0}_{+}}\A, \quad
\rho_{1}=\half\sum_{\G\in\DE^{1}_{+}}\G .
\en
 The Cartan part currents $H^{i}(z)$ are given by
\eq
H^{i}(z)=-\I\A_{+}\pa\vp^{i}
         +\sum_{\A\in\DE^{0}_{+}}\A^{i}:\G_{\A}\B_{\A}:
         +\sum_{\G\in\DE^{1}_{+}}\G^{i}:\xi_{\G}\eta_{\G}:.
\en
After improvement of the energy momentum tensor, the ghost systems acquire
conformal dimensions, $(0,1)$ for $(\beta_\theta,\gamma_\theta )$ and
$(\half ,\half)$ for $(\eta_\gamma ,\xi_\gamma )$.
In the BRST gauge fixing procedure it is natural to assume that the
multiplets $(\B_{\T}, \G_{\T}, b_{\T}, c_{\T})$ and
$(\eta_{\G}, \xi_{\G}, \tilde{b}_{\G}, \tilde{c}_{\G})$
for $\G\in\DE^{1}_{+}$ form  Kugo-Ojima quartets \cite{KuOj}, and that the
corresponding energy-momentum tensor is BRST-exact.
In the case of the $\widehat{sl(N)}$ affine Lie algebra, this has been
proven both  by  homological techniques \cite{Figo}, and by a more direct
method \cite{Haya}.
This ansatz enables us to write the total energy-momentum tensor
in the form:
\eq
T_{total}(z)=T_{ESA}(z)+\{ Q_{BRST}, *\},
\en
where $T_{ESA}$ is the energy-momentum tensor of $G$ extended
superconformal algebra:
\eq
T_{ESA}(z)
= -\half :(\pa\vp)^{2}:
    -\I \left({\rho \over \A_{+}}-\A_{+}\mu\right)\cdot\pa^{2}\vp
    +\sum_{\A\in\DE_{+}(G)}:\B_{\A}\pa\G_{\A}:
 +\half \sum_{\G\in\DE^{1}_{+}}:(\pa\chi_{\G})\chi_{\T-\G}: .
\label{eq:ene}
\en
The central charge of $T_{ESA}$ is given by the formula
\eq
c=n-12 \left({\rho \over \A_{+}}-\A_{+}\mu\right)^{2}
  +2|\DE_{+}(G)|+\half |\DE^{1}_{+}|.
\label{eq:ccemg}
\en
This may be shown to be equal to $c_{total}$ (\ref{eq:cen}).

In the case that $\WG$ is expressed as a direct sum of simple
affine Lie algebras $\oplus_{i} \WG_{i}$,
the relation between the level $k$ of the affine Lie superalgebra $\HBG$
and that, $K_{i}$, of the $i$'th affine Lie algebra $\widehat{G_{i}}$, is
given by considering the decomposition of the vector
$\rho/\A_{+}-\A_{+}\mu$ into the roots of the even subalgebras
$(\oplus_{i}G_{i} )\oplus A_{1}$:
\eq
{\rho \over \A_{+}}-\A_{+}\mu
=\sum_{i}{\rho_{G_{i}}\over \A_{+}}
 +({\half-{|\DE^{1}_{+}|\over 4}\over \A_{+}}-{\A_{+}\over\T^{2}})\T.
\en
Here $\rho_{G_{i}}$ is half the sum of positive roots of the
subalgebra $G_{i}$.
The above decomposition means that
\eq
k+h^{\vee}={\A_{L i}^{2}\over 2}(K_{i}+h^{\vee}_{i}),
\label{eq:levels}
\en
where $h^{\vee}_{i}$ is the dual Coxeter number of the even
subalgebra $G_{i}$ and $\A_{L i}$ is a long root of the subalgebra
$G_{i}$.

The $n$ bosons $\vp(z)=(\vp_{1}(z), \ldots, \vp_{n}(z))$ coupled to the
world sheet curvature, can be divided into two classes, due to the
decomposition of the Cartan subalgebra $\Bh=H_{G}\oplus H_{A_{1}}$,
where $H_{G}$ and $H_{A_{1}}$ are  the Cartan subalgebras of the even
subalgebras $G$ and $A_{1}$, respectively.
A boson $\theta\cdot\vp$ in the $\theta$ direction of the root
space of the even subalgebra $A_{1}$, commutes with the bosons lying
along the root space of the even subalgebra $G$.
The remaining $n-1$ free bosons are used for the free field representation
of the $\hat{G}$ affine Lie algebra, combined with
($\B_{\A}$, $\G_{\A}$)-systems ($\A\in\DE_{+}(G)$).
If we define a Feigin-Fuchs boson $\phi$:
\eq
\phi (z)={\T\cdot\vp(z)\over \sqrt{\T^{2}}},
\en
the energy-momentum tensor (\ref{eq:ene}) becomes
$T_{ESA}(z)=T_{\phi}(z)+T_{G}(z)+T_{\chi}(z)$, where
\eq
T_{\phi}=-\half (\pa\phi)^{2}-{\I Q\over 2} \pa^{2}\phi,
\quad
Q=\sqrt{\T^{2}}
({2-|\DE^{1}_{+}|\over 2\A_{+}}-{2\A_{+}\over\T^{2}}).
\en
$T_{G}(z)$ is the Sugawara energy-momentum tensor of the
affine Lie algebra $\WG$.

So far we have discussed the energy-momentum tensor at the quantum
level using the free field representation.
In the following we will give some examples of the free field
representation of $G$ extended superconformal algebras.
Several of those have been provided previously in the literature.
This is true for $osp(N|2)$ and for $A(n|1)$, $n>1$,
\cite{Mat}, and for $A(1|1)$, \cite{Mats}.
We repeat the results here from our
point of view, partly for illustration and completeness, partly because we
shall need them for the construction of screening operators in sect. 5.
Our results for $D(2|n)$, however, are new. The same is true for our
results for $D(2|1;\alpha)$ presented in ref. \cite{ItMaPe}.

\sect{Examples: Non-exceptional extended superconformal algebras}
In this section we discuss the free field representation of specific
extended superconformal algebras.
The free field representations of $so(N)$ and $u(N)$ extended
superconformal algebras were found in ref. \cite{Mat} from
the viewpoint of the $osp(N|2)$ KdV equations \cite{Kup}.
In ref. \cite{ItMa} two of us gave the classical free field representation
by connecting the DS-gauge and the diagonal gauge
(the Miura transformation).
In the quantum case, we need some modification of the coefficients due
to double contractions in the operator product expansions.
In the present paper we treat the non-exceptional type Lie superalgebras
$A(n|1)$, $B(n|1)$, $D(n|1)$ and $D(2|n)$.
An example of an exceptional type Lie superalgebra is $D(2|1;\A)$,
from which one gets the doubly extended superconformal algebra
$\tilde{\cal A}_{\G}$.
This free field representation is treated in a separate paper
\cite{ItMaPe}.
Although we may treat the other exceptional type Lie superalgebras
$F(4)$ and $G(3)$ in a similar way, we defer explicit
formulas to a separate publication.

\subsect{$A(n|1)$}
Let $\hJ_{i j}$ ($i,j=1,\ldots, N=n+1$) be the $\widehat{sl(N)}$
currents at level $K$, satisfying
\eq
\hJ_{i j}(z)\hJ_{k l}(w)=
\ord{2}{ K(\D_{j k}\D_{i l}-{\D_{i j}\D_{k l}\over N})}
+\ordo{-\D_{j k}\hJ_{i l}(w)+\D_{l i}\hJ_{k j}(w)}+\cdots.
\en
The generators of the $u(N)$ extended superconformal algebras are
the $u(N)$ currents $\hJ_{i j}(z)$, and $2N$ supercurrents
$G_{i}(z)$, $\bG_{i}(z)$ as well as the energy-momentum tensor $T(z)$.
Their free field representations \cite{Mat} are given by
\eqn
J_{i j} &=&\hJ_{i j} -\chi_{i}\bchi_{j}
            +\sqrt{{2\over N(N-2)}}\A_{+}\D_{i j}\pa\hphi , \CR
G_{i} &=& \I\pa\phi\chi_{i}-Q\pa\chi_{i}
            -{\I\sqrt{2}\over \A_{+}}
             (\hJ_{i k}-\chi_{i}\bchi_{k}
              -\sqrt{{N-2\over 2N}}\A_{+}\pa\hphi\D_{i k} )\chi_{k}, \CR
\bG_{i} &=&\I\pa\phi\bchi_{i}-Q\pa\bchi_{i}
            +{\sqrt{2}\I\over \A_{+}}
             \bchi_{k}(\hJ_{k i}-\chi_{k}\bchi_{i}
             -\sqrt{{N-2\over 2N}}\A_{+}\pa\hphi\D_{i k}), \\
T &=&-\half (\pa\phi)^{2}-{\I Q\over 2}\pa^{2}\phi
      -\half (\bchi_{i}\pa\chi_{i}+\chi_{i}\pa\bchi_{i})
      -\half (\pa\hphi)^{2}
      +{1\over 2(K+N)}: \hJ_{i j}\hJ_{j i} : , \nonumber
\enn
where $\A_{+}=\sqrt{K+N}$ and
$Q={\I \sqrt{2}(1-N+\A_{+}^{2})\over \A_{+}}$.
$\hphi(z)$ is a free boson, which represents the $u(1)$ direction of the
even subalgebra, defined as $\hphi(z)=\nu\cdot\vp/\sqrt{\nu^{2}}$ with
$\nu={1\over N-2}\{2(\sum_{i=1}^{N}e_{i})-N(\D_{1}+\D_{2})\}$. Vectors,
$e_i,\D_{i}$ are introduced in Appendix A.
The $u(1)$ current $U(z)$ is given by the trace of a matrix
$J=(J_{i j})$:
\eq
U =J_{i i} =\bchi_{i}\chi_{i}+\sqrt{{2N\over N-2}}\A_{+}\pa\hphi.
\en
One finds that $J_{i j}(z)$ satisfies the operator product
expansions:
\eq
J_{i j}(z)J_{k l}(w)=
\ord{2}{ (K+1)\D_{j k}\D_{i l}-{(K+2)\D_{i j}\D_{k l}\over N-2}}
+\ordo{-\D_{j k}J_{i l}(w)+\D_{l i}J_{k j}(w)}+\cdots.
\en
The supercurrents $G_{i}(z)$ and $\bG_{i}(z)$ belong to the
$N$-dimensional representation of $u(N)$ Kac-Moody algebra
and its conjugate representation, respectively:
\eq
J_{i j}(z)G_{k}(w)=\ordo{-\D_{j k}G_{i}(w)}+\cdots, \quad
J_{i j}(z)\bG_{k}(w)=\ordo{\D_{i k}\bG_{j}(w)}+\cdots .
\en
The nontrivial operator product expansions for the supercurrents are
\eq
G_{i}(z)\bG_{j}(w)
=\ord{3}{{\D_{i j}2(2K+N+2)\over K+N}}
+\ord{2}{-{2(2K+N+2)\over K+N}J_{i j}(w) +\D_{i j}{2(K+2)\over K+N}U(w)}
+\ordo{O_{i j}(w)}+\cdots,
\en
where
\eqn
O_{i j}
&=& \D_{i j}2 T
 -{2K+2+N \over K+N}\pa J_{i j} +\D_{i j}{K+2\over K+N}\pa U  \\
& &+{1\over k+N} \left\{ 2(J_{i k}J_{k j})_{S}-: U J_{i j} :
 +\D_{i j} ( -\half :J_{k l}J_{l k}: +\half :U^{2}:) \right\}.\nonumber
\enn
where we define the symmetric normal ordering $(AB)_{S}(z)$
of two fields $A(z)$ and $B(w)$, by $\half (:AB:+:BA:)(z)$.
Note that for $A(1|1)$ one gets $N=4$ $sl(2)$ extended superconformal
algebra since the $u(1)$ current decouples from the algebra.
Recently the free field representation has been developed in
ref. \cite{Mats}.

\subsect{$osp(N|2)$}
In the definition of the Lie superalgebra $osp(N|2)$, we use a
negative metric for the root space of the even subalgebra $so(N)$,
which makes the level of the affine $\widehat{so(N)}$ negative
(cf. Appendix A).
In order to discuss the representation theory, it is convenient to
change the sign of the metric.

The free field representation of $so(N)$ extended superconformal
algebras, was first given in ref. \cite{Mat}.
Following this, it is convenient to introduce the anti-symmetric
basis for the $so(N)$ currents $\hJ_{i j}(z)$
($1\leq i,j \leq N$) at level $K$, satisfying $\hJ_{i j}=-\hJ_{j i}$.
The operator product expansions are
\eqn
\hJ_{i j}(z)\hJ_{l m}(w)&=&
\ord{2}{-K(\D_{i l}\D_{j m}-\D_{i m}\D_{j l})} \CR
& & +\ordo{-\D_{i l}\hJ_{j m}(w)+\D_{j l}\hJ_{i m}(w)+\D_{j m}\hJ_{l i}(w)
       -\D_{i m}\hJ_{l j}(w)}+\cdots .
\enn
Let $\psi_{i}$ ($i=1,\ldots, N$) be real fermions satisfying
$ \psi_{i}(z)\psi_{j}(w)={\D_{i j}\over z-w}+\cdots$.
The generators of the $so(N)$ extended superconformal algebras
are given by
\eqn
T &=&-\half (\pa \phi)^{2}-{\I Q \over 2}\pa^{2}\phi
      -\half :\psi_{i}\ppsi_{i}:
      -{1\over 2(K+N-2)} :\hJ_{i j}\hJ_{j i}:, \CR
G_{i} &=&\I\pa \phi\psi_{i} - Q\ppsi_{i}
       +{\I\over \A_{+}}\hJ_{i j}\psi_{j}, \CR
J_{i j} &=& \hJ_{i j}+\psi_{i}\psi_{j},
\enn
where $\A_{+}=\sqrt{K+N-2}$ and $Q={\I(2-N-\A_{+}^{2})\over \A_{+}}$.
The currents $J_{i j}$ satisfy the level $K+1$ $\widehat{so(N)}$
Kac-Moody algebra.
Other nontrivial operator product expansions of generators are
\eqn
J_{i j}(z)G_{k}(w)&=&\ordo{\D_{j k} G_{i}(w)-\D_{i k}G_{j}(w)}+\cdots, \CR
G_{i}(z)G_{j}(w)
&=&\ord{3}{{2K^{2}+2K+N-2 \over K+N-2} \D_{i j}}
 +\ord{2}{{2K+N-2\over K+N-2}J_{i j}(w)}
+\ordo{O_{i j}(w)}+\cdots,
\enn
where
\eq
O_{i j}= \D_{i j} 2 T+\half {2K+N-2\over K+N-2}\pa J_{i j}
-{1\over K+N-2}\left\{ (J_{i k} J_{k j})_{S}
  -\D_{i j}:J_{l m}J_{m l}: \right\}.
\en
\subsect{$D(2|n)$}
The Lie superalgebra $D(2|n)$ has the even subalgebra $so(4)\oplus sp(2n)$.
Since $so(4)$ is isomorphic to $sl(2)\oplus sl(2)$, we can use the
hamiltonian reduction to find an $sl(2)\oplus sp(2n)$ extended
superconformal algebra.
In this case we use a negative metric for the root space of the $sl(2)$
subalgebra, and a positive metric for the root space of the $sp(2n)$
subalgebra, so the levels of the $sl(2)$ and of the $sp(2n)$ algebras have
opposite signs.
Let $\hat{J}$ and $\hat{I}$ be the $sp(2n)$ currents
with level $K$ and the $sl(2)$ currents with level $\tilde{K}$.
{}From (\ref{eq:levels}) we find that the level $\tilde{K}$ is equal to
$-K-2n-2$.
An element of $sp(2n)$ may be represented by a matrix of the form
\eq
M = \left ( \begin{array}{cc} A & B \\ C & -\R A \end{array} \right ),
\en
where $A$, $B$ and $C$ are $n \times n$ matrices, the matrix $\R A$ is
defined as $\R A_{i,j} = A_{n+1-j,n+1-i}$, and $B = \R B$, $C = \R C$.
We denote the currents in the $sp(2n)$ matrix
\mbox{\bf $\hat{J}$} as ($i,j \leq n$)
\eqn
\hat{J}_{ij} & = & \hat{J}_{-i+j},                 \CR
\hat{J}_{2n+1-i,j} & = & \hat{J}_{+i+j},           \CR
\hat{J}_{i,2n+1-j} & = & \hat{J}_{-i-j}.
\enn
The Cartan algebra element $H^i = \hat{J}_{ii}$ is written
as $\hat{J}_{+i-i}$ in this notation.
Note that
$\hat{J}_{+i+j} = \hat{J}_{+j+i}$ and
$\hat{J}_{-i-j} = \hat{J}_{-j-i}$.
We write the $sl(2)$ currents
as $\hat{I}^{\pm}$ and $\hat{I}^{3}$.
Introduce $4n$ free fermions $\psi^{+}_{\pm i}$ and $\psi^{-}_{\pm i}$
with the OPE's
$\psi^{-}_{\pm i}(z)\psi^{+}_{\mp j}(w)= \zwon{\D_{ij}}+\cdots$.
We can then write the $sp(2n)\oplus sl(2)$ Kac-Moody currents of the
extended superconformal algebra as
\eq
I^{3}  =  \hat{I}^{3} - \half :\psi^{-}_{+m}\psi^{+}_{-m}:
- \half :\psi^{-}_{-m} \psi^{+}_{+m}:, \quad
I^{\pm}  =  \hat{I}^{\pm} + \psi^{\pm}_{+m}\psi^{\pm}_{-m},
\en
and
\eqn
J_{+i-j} & = & \hat{J}_{+i-j} + :\psi^{+}_{+i}\psi^{-}_{-j}:
+ :\psi^{-}_{+i}\psi^{+}_{-j}:,        \CR
J_{+i+j} & = & \hat{J}_{+i+j} + \psi^{+}_{+i}\psi^{-}_{+j}
- \psi^{-}_{+i}\psi^{+}_{+j},        \CR
J_{-i-j} & = & \hat{J}_{-i-j} + \psi^{+}_{-i}\psi^{-}_{-j}
- \psi^{-}_{-i}\psi^{+}_{-j},
\enn
and the free field representation of the supercurrents are given by
\eqn
G^{\pm}_{+i} = -\frac{Q}{\sqrt{2}} \pa\psi^{\pm}_{+i}
+ \frac{\I}{\sqrt{2}} \psi^{\pm}_{+i}\pa\phi
+ \frac{\I}{\A_{+}} \left ( \mp :\psi^{\pm}_{-m}J_{+m+i}:
- :\psi^{\pm}_{+m} J_{+i-m}: \pm \psi^{\pm}_{+i} \hat{I}^{3}
+ \psi^\mp_{+i}\hat{I}^{\pm} \right ),   \CR
G^{\pm}_{-i} = -\frac{Q}{\sqrt{2}} \pa\psi^{\pm}_{-i}
+ \frac{\I}{\sqrt{2}} \psi^{\pm}_{-i}\pa\phi
+ \frac{\I}{\A_{+}} \left ( \mp :\psi_{+m}J_{-i-m}:
+ :\psi^{\pm}_{-m} J_{+m-i}: \pm \psi^{\pm}_{-i} \hat{I}^{3}
- \psi^\mp_{-i}\hat{I}^{\pm}
\right ),  \CR
\enn
where $\A_{+} = \sqrt{K+2n}$ and
$Q = -\I\sqrt{2} \frac{\A_{+}^2 - 2n + 1}{\A_{+}}$.
The superscript $+$ corresponds to an $sl(2)$ isospin of $+\half$ for the
supercurrents and fermions, and $-$ corresponds to isospin $-\half$,
while the subscripts $\pm i$ corresponds to a numbering of the weights
of the $2n$-dimensional vector representation of $sp(2n)$.
With these representations, we get the standard OPE's for the Kac-Moody
currents, but with levels $K+2$ for $sp(2n)$ and $-K-n-2$
for $sl(2)$:
\eqn
J_{+i-j}(z)J_{+k-l}(w) & = & \zwtw{(K+2)\D_{il}\D_{jk}}
+ \zwon{\D_{kj}J_{+i-l}(w) - \D_{il}J_{+k-j}(w)}+\cdots,   \CR
J_{+i-j}(z)J_{+k+l}(w) & = &
\zwon{\D_{jl}J_{+i+k}(w) + \D_{jk}J_{+i+l}(w)}+\cdots,     \CR
J_{+i-j}(z)J_{-k-l}(w) & = &
-\zwon{\D_{ik}J_{-j-l}(w) + \D_{il}J_{-j-k}(w)}+\cdots,    \CR
J_{-i-j}(z)J_{+k+l}(w) & = &
\zwtw{(K+2)(\D_{il}\D_{kj}+\D_{ik}\D_{jl})}                \CR
&-& \!\!\!\!
 \zwon{\D_{kj}J_{l-i}(w) + \D_{ik}J_{l-j}(w) + \D_{il}J_{k-j}(w)
+ \D_{jl}J_{k-i}(w)}+\cdots,
\enn
\eqn
I^{3}(z) I^{\pm}(w) & = & \zwon{\pm I^{\pm}(w)}+\cdots,  \quad\quad
I^{3}(z) I^{3}(w)~~ =~~ \zwtw{-(K+n+2)/2}+\cdots,              \CR
I^{+}(z) I^{-}(w) & = & \zwtw{-K-n-2} + \zwon{2I^{3}(w)}+\cdots.
\enn
The OPE's of the $sp(2n)$ Kac-Moody currents with the supercurrents $G$ are
\eqn
\label{opeJG}
J_{-i+j}(z)G^{\pm}_{+k}(w) & = &
\zwon{\D_{ik}G^{\pm}_{+j}(w)}+\cdots,                        \CR
J_{-i+j}(z)G^{\pm}_{-k}(w) & = &
\zwon{-\D_{jk}G^{\pm}_{-i}(w)}+\cdots,                       \CR
J_{+i+j}(z)G^{\pm}_{-k}(w) & = &
\zwon{\pm(\D_{ik}G^{\pm}_{+j}(w) + \D_{jk}G^{\pm}_{+i}(w))}+\cdots, \CR
J_{-i-j}(z)G^{\pm}_{+k}(w) & = &
\zwon{\pm(\D_{ik}G^{\pm}_{-j}(w) + \D_{jk}G^{\pm}_{-i}(w))}+\cdots,
\enn
and the OPE's of the $sl(2)$ Kac-Moody currents with the
supercurrents are
\eqn
\label{opeIG}
I^{3}(z) G^{-}_{\pm i}(w) & = & \zwon{-\half G_{\pm i}(w)}+\cdots,
\quad \quad \quad
I^{3}(z) G^{+}_{\pm i}(w) ~~=~~ \zwon{\half G_{\pm i}(w)}+\cdots, \CR
I^{-}(z) G^{+}_{\pm i}(w) & = & \zwon{\pm G^{+}_{\pm i}(w)}+\cdots,
\quad \quad \quad
I^{+}(z) G^{-}_{\pm i}(w)~~ =~~ \zwon{\pm G^{-}_{\pm i}(w)}+\cdots.
\enn
Finally we get the slightly more complicated expression for the
operator product of two supercurrents
\eqn
G^{\pm}_{+i}(z) G^{\pm}_{+j}(w)
& = & \zwon{\pm {2\over \A_{+}^{2}}J_{+i+j} I^{\pm}(w)}+\cdots, \CR
G^{\pm}_{-i}(z) G^{\pm}_{-j}(w)
& = & \zwon{\pm {2\over \A_{+}^{2}}J_{-i-j} I^{\pm}(w)}+\cdots, \CR
G^{\pm}_{-i}(z) G^{\pm}_{+j}(w)
&  = &  \zwtw{-{2\over\A_{+}^{2}}\D_{i j}K I^{\pm}(w)}
+ \zwon{-{1\over\A_{+}^{2}} \D_{ij}K\pa I^{\pm}(w)
         + {2\over\A_{+}^{2}} J_{-i+j} I^{\pm}(w)}
+\cdots, \CR
G^{-}_{\pm i}(z)G^{+}_{\pm j}(w)
& = &  \zwtw{{2(K+n+2)\over \A_{+}^{2}}J_{\pm(i+j)}(w)}
     + \zwon{O^{-+}_{\pm i\pm j}(w)} + \cdots                           \CR
G^{-}_{\pm i}(z)G^{+}_{\mp j}(w)
& = &  \zwth{{\D_{ij}((K+1)(K+2n)+K)\over \A_{+}^{2}}}               \CR
& & +\zwtw{\pm {2(K+n+2)\over \A_{+}^{2}}J_{\pm(i-j)}(w)
            - {2K\over \A_{+}^{2}}\D_{ij}I^{3}(w)}
 + \zwon{O^{-+}_{\pm i\mp j}(w)} + \cdots,
\enn
where
\eqn
O^{-+}_{\pm i\pm j} &=&
{K+n+2\over\A_{+}^{2}}\pa J_{\pm(i+j)}
            + {2\over \A_{+}^{2}}J_{\pm(i+j)} I^{3}    \CR
 & &  +{1\over \A_{+}^{2} }(J_{k\pm i} J_{\pm j-k})_{S}
-{1\over \A_{+}^{2}} (J_{\pm i-k}J_{k\pm j})_{S},  \CR
 O^{-+}_{\pm i\mp j} &=&
\pm\mbox{[}{(K+n+2)\over \A_{+}^{2}} \pa J_{\pm(i-j)}
                +{2\over\A_{+}^{2}} J_{\pm(i-j)}I^{3} \mbox{]}  \CR
& &
+{1\over\A_{+}^{2}}\mbox{[} (J_{\pm(i-k)}J_{\pm(k-j)})_{S}
        + (J_{\pm(i+k)}J_{\mp(k+j)})_{S} \mbox{]} \\
& & + \D_{ij}\mbox{[}- {1\over 4\A_{+}^{2}} \tr :J^2:
 +{1\over \A_{+}^{2}}\{K\pa I^{3}
 + 2 :(I^{3})^2:+ 2 (I^{-}I^{+})_{S} \}
 + T \mbox{]}. \nonumber
\enn
Here the energy momentum tensor $T$ is
$T = T_{\phi} + T_{sl(2)} + T_{sp(2n)} + T_\psi$, and
\eqn
T_{\phi} & = & -\frac{1}{2}:(\pa\phi)^2:
-\frac{\I Q}{2} \pa^2\phi,                           \CR
T_{sl(2)} & = & - (\A_{+})^{-2} \left ( (\hat{I}^{3})^2
+ \hat{I}^{-}\hat{I}^{+} \right )_{S},                       \CR
T_{sp(2n)} & = & (\A_{+})^{-2} \frac{1}{4} \tr :\hat{J}^2:,  \CR
T_\psi & = & \half \left (
:\pa \psi^{-}_{+m} \psi^{+}_{-m}: + :\pa \psi^{-}_{-m} \psi^{+}_{+m}:
+:\pa \psi^{+}_{+m} \psi^{-}_{-m}: + :\pa \psi^{+}_{-m} \psi^{-}_{+m}:
\right ).
\enn
These OPE's are identical with the recent results obtained by a different
approach in ref.\cite{BoFrLi}.
In the next section we shall discuss the structure of screening operators
for $G$ extended superconformal algebras.

\sect{Degenerate Representation of $G$ extended superconformal algebras}
In this section we  discuss the representation theory of an
extended superconformal algebra with $\WG$ affine Lie algebra
symmetry using free fields.
The Fock space of the $G$ extended superconformal algebras is a tensor
product of ones for
$|\DE^{1}_{+}|$ fermions $\chi_{\G}$, free fields for the affine Lie
algebra $\WG$, and a free boson $\phi(z)$ coupled to the world
sheet curvature.

The free field representations of affine Lie algebras $\WG$ are
studied in refs. \cite{FeFr}.
Let $\bA_{i}$ ($i=1,\ldots, r$) be simple roots of the even subalgebra $G$.
$\blm_{i}$ ($i=1, \ldots, r$) the fundamental weights of $G$ satisfying
${2\blm_{i}\cdot\bA_{j} \over \bA_{j}^{2}}=\D_{i j}$.
$\Phi^{\bLM}_{\blm}(z)$ is a primary field of the affine Lie algebra
$\widehat{G}$ at level $K$, with weight $\blm$ in the highest
weight module with highest weight $\bLM$.
In the free field representation, this field can be expressed
as $p^{\bLM}_{\blm}(z)\EXP ({\I\bLM\cdot\vp(z)\over\A_{+}})$, where
$p^{\bLM}_{\blm}$ is a polynomial consisting of terms,
$\G_{\A_{1}}\cdots \G_{\A_{k}}(z)$ ($\A_{i}\in\DE_{+}(G)$)
such that $\blm=-\bLM+\A_{1}+\cdots+\A_{k}$.
Note that in the present prescription, the vertex operator
$\EXP ({\I\bLM\cdot\vp(z)\over\A_{+}})$ represents the lowest weight state
$\Phi^{\bLM}_{-\bLM}(z)$.

Denote the total Fock space as
$F_{\chi,\bLM,p}=F^{\chi}\otimes F^{G}_{\bLM}\otimes F^{\phi}_{p}$,
where $F^{\chi}$ is a fermionic Fock space constructed from $\chi_{\G}$
($\G\in\DE^{1}_{+}$).
$F^{G}_{\bLM}$ is a Fock space of the algebra $\widehat{G}$ built on a
primary field $\Phi^{\bLM}_{-\bLM}=\E^{{\I\bLM\cdot\vp(z)\over \A_{+}}}$.
$F^{\phi}_{p}$ is a Fock space built on a vertex operator
$V_{p}(z)=\E^{\I p\sqrt{\T^{2}}\phi(z)}$.
The dual spaces $(F^{G}_{\bLM})^{*}$ and $(F^{\phi}_{p})^{*}$ are isomorphic
to $F^{G}_{-2\rho_{G}-\bLM}$ and $F^{\phi}_{-Q-p}$, respectively.

A primary field of a $G$ extended superconformal algebra is expressed
as the products of three fields:
\eq
V_{\G_{1}, \ldots, \G_{l}}{}^{\LM}_{\lm}{}_{p}(z)
=\chi_{\G_{1}}(z)\cdots\chi_{\G_{l}}(z)\Phi^{\LM}_{\lm}(z)
 \E^{\I p\sqrt{\T^{2}}\phi(z)},
\label{eq:prim}
\en
where $\G_{i}$ are positive odd roots.
The conformal weight of (\ref{eq:prim}) is given by
\eq
\DE={l\over2}+{\LM(\LM+2\rho_{G})\over 2\A_{+}^{2}}
+\half (p^{2}+Q p)\T^{2}.
\en

\subsect{Screening operators}
In order to study the representation of the algebra using free fields,
we must specify the screening operators which commute with the generators
of the extended superconformal algebra.
We consider screening operators which correspond to the simple
roots of the Lie superalgebra $\BG$.
These screening operators are BRST-equivalent to those of the
affine Lie superalgebra $\HBG$ \cite{BeOo}.
In the present choice of the simple root system of the Lie superalgebra in
table \ref{ta:li}, the simple roots of the even subalgebra $G$ are a
subset of those of $\BG$ (see Appendix A).
Thus we will get the screening operators corresponding to
the simple roots of the affine Lie algebra $\WG$.
Since the remaining simple roots are odd,
we will obtain  fermionic type screening operators.
In addition, we will find another screening operator, which is
necessary to characterize the $A_{1}$ even subalgebra corresponding
to the root $\T$.
\subsubsection{Affine screening operators.}
First we can take the screening operators $S_{\bA_{i}}(z)$ of the affine
Lie algebra $\widehat{G}$ as those of $G$ extended superconformal algebra:
\eq
S_{\bA_{i}}(z)=s_{\bA_{i}}(z)\E^{-{\I\bA_{i}\cdot\vp(z)\over\A_{+}}},
\label{eq:scr1}
\en
where $s_{\bA_{i}}$ consists of terms like $\B_{\bA_{i}}$  and
$\G_{\A_{1}}\cdots \G_{\A_{k}}\B_{\A_{1}+\cdots+\A_{k}+\bA_{i}}$
with $\A_{1}, \ldots, \A_{k}\in\DE_{+}(G)$.
These screening operators are used for the characterization of
singular vectors in the Fock modules of the affine Lie algebra
$\WG$.

\subsubsection{ Fermionic screening operators.}
Next, we consider a class of screening operators, which
corresponds to the odd simple roots.
In the diagonal gauge (free field realization)
we have fermions for every negative odd root.
We expect that the resulting screening operators are expressed
as some linear combinations of the odd root fermions $\chi_{\G}$.
Remember that the negative odd root space $(\BG_{1})_{-\half}$
belong to the fundamental representation of the even subalgebra
$G$ of dimensions $|\DE^{1}_{+}|$.
Denote the highest weight of these representations as
$\bLM^{*}$.
To each odd positive root corresponds a weight vector in the
representation with the highest weight $\bLM^{*}$.
Let $\G(\blm)$ be an odd positive root associate with the weight
vector $\blm$.
In table \ref{ta:we} we list such highest weights for
every $\BG$.
In order that this type of screening operator commutes with the
$\WG$ currents, this should be a $\WG$ singlet operator {\it i.e.}
the operator product expansion with the $\WG$ currents should be
regular.
These observations lead to fermionic screening operators of
the following type:
\eq
S_{f}(z)= \sum_{\blm}\chi_{\G(\blm)}\Phi^{\bLM^{*}}_{\blm}
          \EXP(-{\I\sqrt{\T^{2}}\phi(z)\over 2\A_{+}}),
\label{eq:scr2}
\en
where $\blm$ runs over the weights of the representation with
the highest weight $\bLM^{*}$.

For example, in the case of $A(n|1)$, there are two highest weight
vectors  $\blm_{1}\oplus \nu$ and $\blm_{n}\oplus (-\nu)$.
The latter weight vector is the conjugate representation of the
former.
$\nu$ denotes the $u(1)$ charge.
We find two fermionic screening operators
\eqn
S_{f}(z)&=& \chi_{i}\Phi_{i}
\E^{-\sqrt{{N\over N-2}}\A_{+}\hphi(z)+{\phi(z)\over\sqrt{2}\A_{+}}},\CR
\bar{S}_{f}(z)&=& \bchi_{i}\bPhi_{i}
\E^{\sqrt{{N\over N-2}}\A_{+}\hphi(z)+{\phi(z)\over\sqrt{2}\A_{+}}},
\label{eq:scr3}
\enn
where $\Phi_{i}(z)$ and $\bPhi_{i}(z)$ ($i=1, \ldots ,N$) are primary
fields of $\hJ_{i j}$  satisfying
\eqn
J_{i j}(z)\Phi_{k}(z)&=&
\ordo{\D_{i k}\Phi_{j}(w)-{1\over N}\D_{i j}\Phi_{k}(w)}+\cdots, \CR
J_{i j}(z)\bPhi_{k}(z)&=&
\ordo{-\D_{k j}\bPhi_{i}(w)+{1\over N}\D_{i j}\bPhi_{k}(w)}+\cdots.
\label{eq:currscr1}
\enn
We can easily check that $S_{f}(z)$ and $\bar{S}_{f}(z)$ are $u(N)$ singlets
and have conformal dimension one.
Moreover $S_{f}(z)$ satisfy
\eq
G_{i}(z)S_{f}(w)=\mbox{regular}, \quad
\bG_{i}(z)S_{f}(w)={\pa\over \pa w}
\left( \ordo{\I\sqrt{2}\A_{+}\Phi_{i}
\E^{-\sqrt{{N\over N-2}}\A_{+}\hphi(w)+{\phi(w)\over\sqrt{2}\A_{+}}}}
\right)+\cdots,
\en
by virtue of the Knizhnik-Zamolodchikov equations \cite{KnZa}:
\eq
\pa \Phi_{i}={1\over K+N}:J_{k i}\Phi_{k}: .
\en
Similar relations hold for $\bar{S}_{f}(z)$.

In the case of $so(N)$ extended superconfomal algebras, we
find the following fermionic screening operator:
\eq
S_{f}(z)=\psi_{i}(z)\Phi_{i}(z)\E^{-{1\over\A_{+}}\phi (z)},
\label{eq:scr4}
\en
where $\Phi_{i}(z)$ belongs to the $N$ dimensional representation of
$\widehat{so(N)}$:
\eq
\hJ_{i j}(z)\Phi_{k}(w)=\ordo{\D_{j k}\Phi_{i}(w)-\D_{i k}\Phi_{j}(w)}
+\cdots .
\en
$S_{f}(z)$ is an $so(N)$ singlet operator and has conformal dimension
one.
The operator product expansions with the supercurrents are
\eq
G_{i}(z)S_{f}(w)={\pa\over\pa w}
\left(\ordo{\I\A_{+}\Phi_{i}(w)\E^{-{1\over\A_{+}}\phi (w)}} \right)
+\cdots,
\en
We can now also construct fermionic screening operators for the
$D(2|n)$ case.
Let $\Phi^{\pm}_{\pm k}$ be primary fields in the $2n\times 2$
dimensional representation of $sp(2n)\oplus sl(2)$, which transform as
$G$ under $\hat{J}$ and $\hat{I}$, i.e. we have
\eqn
J_{-i+j}(z)\Phi^\pm_{+k}(w) & = &
\zwon{\D_{ik}\Phi^\pm_{+j}(w)}+\ldots, \quad \quad \quad
J_{-i+j}(z)\Phi^\pm_{-k}(w)~~=~~
\zwon{-\D_{jk}\Phi^\pm_{-i}(w)}+\ldots,                             \CR
J_{+i+j}(z)\Phi^\pm_{-k}(w) & = &
\zwon{\pm(\D_{ik}\Phi^\pm_{+j}(w) + \D_{jk}\Phi^\pm_{+i}(w))}+\ldots,\CR
J_{-i-j}(z)\Phi^\pm_{+k}(w) & = &
\zwon{\pm(\D_{ik}\Phi^\pm_{-j}(w) + \D_{jk}\Phi^\pm_{-i}(w))}+\ldots,
\CR
I^3(z) \Phi^-_{\pm i}(w) & = & \zwon{-\half \Phi_{\pm i}(w)}+\ldots,
\quad \quad \quad
I^3(z) \Phi^+_{\pm i}(w) ~~= ~~ \zwon{\half \Phi_{\pm i}(w)}+\ldots,
\CR
I^-(z) \Phi^+_{\pm i}(w) & = & \zwon{\pm \Phi^+_{\pm i}(w)}+\ldots,
\quad \quad \quad
I^+(z) \Phi^-_{\pm i}(w)~~ =~~ \zwon{\pm \Phi^-_{\pm i}(w)}+\ldots.
\enn
We can then write the fermionic screening operator as
\eq
S_f(z) = (\psi^+_{+k}\Phi^-_{-k} + \psi^+_{-k}\Phi^-_{+k}
+ \psi^-_{+k}\Phi^+_{-k} + \psi^-_{-k}\Phi^+_{+k})(z)
e^{-\frac{1}{\sqrt{2}\A_+}\phi(z)}.
\en
The OPE's of the Kac-Moody currents with this screening operator is regular,
and the OPE's with the supercurrents are
\eqn
G^\pm_{+i}(z) S_f(w)
& = & \frac{\pa}{\pa w} \left ( \zwon{\I\A_+ \Phi^\pm_{+i}(w)
e^{-\frac{1}{\sqrt{2}\A_+}\phi(w)}}
\right ) + \cdots,  \CR
G^\pm_{-i}(z) S_f(w)
& = & \frac{\pa}{\pa w} \left ( \zwon{\I\A_+ \Phi^\pm_{-i}(w)
e^{-\frac{1}{\sqrt{2}\A_+}\phi(w)}}
\right ) + \cdots.
\enn
A similar construction is possible for other $G$ extended superconformal
algebras.
\subsubsection{ Screening operators in the $\theta$-direction.}
Finally we need a screening operator which is necessary to characterize
the $\T$ direction.
Denote this screening operator as $S_{\T}(z)$.
We assume that this operator takes the form:
\eq
S_{\T}(z)=s_{\T}(z) \EXP( {2\I\A_{+}\phi(z)\over \sqrt{\T^{2}}}),
\label{eq:scr5}
\en
where $s_{\T}(z)$ is a $G$ singlet operator with conformal dimension
$|\DE^{1}_{+}|/2$ containing a term $\prod_{\G\in\DE^{1}_{+}}\chi_{\G}$.
This assumption is justified in part by the following list of results:

In the case of $N=4$ $sl(2)$ superconformal algebra this type of
screening operator has been obtained in ref. \cite{Mats}.
When we consider the $osp(N|2)$ case, for $N=1$ and $2$, the screening
operators are given as $S_{\T}=\psi_{1}\EXP(-\A_{+}\phi(z))$ and
$S_{\T}=(\psi_{1}\psi_{2}-{1\over K}J_{12})\EXP(-\A_{+}\phi(z))$,
respectively.
These are nothing but the  screening operators of $N=1$ and $2$ minimal
models.
For the $N=3$ case, this kind of screening operator has been found in ref.
\cite{Miki}, the result of which becomes in our notation:
\eq
S_{\T}(z)=
\mbox{[} \psi_{1}\psi_{2}\psi_{3}
-{1\over K}(J_{12}\psi_{3}+J_{32}\psi_{1}+J_{13}\psi_{2})\mbox{]}(z)
\E^{\A_{+}\phi(z)}.
\label{eq:scr6}
\en
We expect that this procedure can be generalized to any $N$.
In particular, we find that for $N=4$, the screening operator
takes the form:
\eqn
S_{\T}(z)&=&
\{ \psi_{1}\psi_{2}\psi_{3}\psi_{4}
-{1\over K}(J_{12}\psi_{3}\psi_{4}-J_{13}\psi_{2}\psi_{4}
            +J_{14}\psi_{2}\psi_{3}-J_{24}\psi_{1}\psi_{3}
             +J_{34}\psi_{1}\psi_{2}+J_{23}\psi_{1}\psi_{4}) \CR
& & +{1\over K(K+2)}
\mbox{[}(J_{12}J_{34})_{S}-(J_{13}J_{24})_{S}
-(J_{14}J_{23})_{S}\mbox{]} \}(z)
\E^{\A_{+}\phi(z)}.
\label{eq:scr7}
\enn
The operator product expansions of the supercurrents with this screening
operator is
\eq
G_1(z)S_\T(w) = \frac{\pa}{\pa w}
\left ( \frac{\I}{\A_+} \zwon{O(w) \EXP(\A_+\psi(w))} \right )
+ \cdots ,
\en
where
\eq
O =
-\psi_2 \psi_3 \psi_4
+ \frac{2}{K(K+2)}(J_{23} \psi_4 +J_{34} \psi_2 +J_{42} \psi_3 ).
\en
Similar expressions hold for the other supercurrents.
For general $N$, we have a conjecture for the form $s_{\T}(z)$.
However due to the complicated operator product expansions
we have not yet succeeded in a complete verification.
As an example of a further nontrivial case,
we have constructed the screening operator
for $D(2|1;\A)$ \cite{ItMaPe}.
For generic $G$ we can easily show that  the field (\ref{eq:scr5}) has
conformal dimension one.
In the following we assume the existence of this kind of screening
operator for any $G$.

\subsection{Structure of null fields}
Based on the above observations on the screening operators, we discuss
the structure of singular vectors of $G$ extended superconformal algebras.
We consider the Neveu-Schwarz sector for simplicity.

Firstly we consider the singular fields corresponding to the affine Lie
algebra $\WG$ \cite{FeFr}.
These are given by the following screened vertex operators:
\eq
\Psi^{\bLM}_{\B_{1},\ldots, \B_{m}}(z)=\oint d u_{1} \cdots d u_{m}
          S_{\B_{1}}(u_{1})\cdots S_{\B_{m}}(u_{m})
          \Phi^{\bLM}_{-\bLM}(z),
\label{eq:singkm}
\en
where $\B_{a}$ ($a=1,\ldots, m$) are simple roots of the algebra $G$:
i.e. $\B_{a}=\bA_{i_{a}}$ for some $i_{a}=1, \ldots, r$.
The contours of integrations are taken as in ref. \cite{KaMa}.
If the above integral exists and is nonzero, this gives the null fields in
the Fock space $F^{G}_{\bLM+\sum_{a=1}^{m}\B_{a}}$.
The condition for existence of the null fields is given by the
\lq\lq on-shell" conditions \cite{KaMa}:
\eq
{1\over\A_{+}^{2}}\sum_{a<b} \B_{a}\cdot\B_{b}
-{1\over\A_{+}^{2}}\sum_{a=1}^{m}\B_{a}\cdot\bLM=-M,
\label{eq:onkm}
\en
with a positive integer $M_{i}$.
If $\sum_{a=1}^{m}\B_{a}=n'\bA$ and $M=n n'$ for a positive root
$\bA\in\DE_{+}(G)$ and positive integers $n$ and $n'$, we get the
Kac-Kazhdan formula \cite{KaKa}:
\eq
(\bLM+\rho_{G})\cdot \bA=-n\A_{+}^{2}+n' {\bA^{2}\over2}.
\label{eq:kaka}
\en
This formula and its dual, which is obtained by replacing $\bLM$
by $-2\rho_{G}-\bLM$, characterize the singular vectors of the
Fock module of the affine Lie algebra $\WG$.

The fermionic singular vectors are given by considering a screened
vertex operator of the form:
\eq
\Psi^{\bLM,p}(z)=\oint d u S_{f}(u)V_{\bLM,p}(z),
\en
where $V_{\bLM,p}(z)=\Phi^{\bLM}_{\blm}\EXP(\I p\sqrt{\T^{2}}\phi(z))$.
The non-zero existence of the above contour integral requires
the condition:
\eq
{\bLM\cdot \bLM^{*} \over \A_{+}^{2}}-{p\T^{2}\over 2\A_{+}}=-M,
\label{eq:fersi}
\en
where $M$ is a positive integer. In this case $\Psi^{\LM,p}(z)$ is
a singular vector in the Fock module
$F^{G}_{\bLM+\bLM^{*}}\otimes F^{\phi}_{p-{1\over 2\A_{+}}}$ at
level $M-\half$.

The null fields in the $\T$ direction can be obtained from the
screened vertex operators:
\eq
\Psi^{p}_{r}(z)=\oint d u_{1}\cdots d u_{r}
              S_{\T}(u_{1})\cdots S_{\T}(u_{r}) V_{p}(z),
\en
where $V_{p}(z)=\E^{\I p\sqrt{\T^{2}}\phi (z)}$.
The on-shell condition becomes
\eq
{r(r-1)\over 2}({2\A_{+}\over \T^{2}})^{2}\T^{2}+2r p \A_{+}=-M.
\en
with an positive integer $M$.
Writing $M$ as ${r (s -|\DE^{1}_{+}|+2) \over 2}$, we find that
$p$ is given by
\eq
p_{r,-s}=-{r-1\over \T^{2}}\A_{+}-{s-|\DE^{1}_{+}|+2 \over 4\A_{+}}.
\en
In this case $\Psi^{p}_{r}(z)$ is a null field in the Fock module
$F^{\phi}_{-Q-p_{r,s}}$ at level ${r s\over 2}$, where
\eq
p_{r,s}={1-r\over \T^{2}}\A_{+}+{s+|\DE^{1}_{+}|-2 \over 4\A_{+}}.
\label{eq:virsi}
\en
The precise range for $s$ cannot be identified unambiguously until the
multiplicities of zeros of the Kac determinant (or something equivalent)
has been dealt with.

The formulas (\ref{eq:kaka}), (\ref{eq:fersi}) and (\ref{eq:virsi})
characterize the whole singular vector structure of the $G$
extended superconformal algebras.

\sect{Conclusions and Discussion}
In the present paper we have studied $G$ extended superconformal
algebras from the viewpoint of  classical and quantum hamiltonian
reductions of an affine Lie superalgebra $\HBG$, with even Lie
subalgebras $\WG\oplus \widehat{sl(2)}$.
At the classical level we have derived generic formulas for the
commutation relations of the ensuing $G$ extended superconformal algebras.
At the quantum level we have obtained the quantum expression of the
energy momentum tensor for general types by adopting free field
representations for the affine Lie superalgebra.
We have given explicit free field expressions for the supercurrents
in the case of the non-exceptional type Lie superalgebras $sl(N|2)$,
$osp(N|2)$ and $osp(4|2N)$.
For the exceptional Lie superalgebra $D(2|1;\A)$, we obtain the
$N=4$ $sl(2)\oplus sl(2)$ doubly extended superconformal algebra
$\tilde{{\cal A}}_{\G}$ ($\G={\A\over 1-\A}$), which admits one continuous
parameter $\A$\cite{dob}.
A detailed analysis of the free field representations is presented
in a separate publication \cite{ItMaPe}.
For the other exceptional type Lie superalgebras $F(4)$ and $G(3)$,
we can construct a Poisson bracket structure of $N=8$ $spin(7)$ and
$N=7$ $G_{2}$ extended superconformal algebras, respectively, by
introducing a pseudo matrix representation\cite{DeNi}.
Although we do not present their free field representation of
supercurrents in this paper, we expect that similar formulas holds as
in the non-exceptional case.
This subject will be treated elsewhere.
We have introduced a set of screening operators for $G$ extended
superconformal algebras.
Using the null field construction,
we have identified the primary fields corresponding to degenerate
representations of these algebras.

In order to make further progress on the representation theory, one should
for example investigate  the Kac determinant formula.

It is now in principle possible to calculate correlation functions and
characters for these models. They will then be expressed as
products of those of $\WG$ affine Lie algebras, Virasoro
minimal models and free fermions.

Compared to the linearly extended superconformal algebras, a
geometrical interpretation of the present non-linear algebras
is unclear.
It seems an interesting problem to try to find a way of interpreting
these non-linear symmetries in terms of non-linear $\sigma$-models on
non-symmetric Riemannian manifolds.
This problem is important in order to clarify the geometrical
meaning of the $W$-algebras.

The present construction of the extended superconformal algebra
is based on the hamiltonian reduction of the affine Lie superalgebras.
It is well understood that in the case of $W$-algebras associated with
simple Lie algebras, the hamiltonian reduction
\cite{DrSo}-\cite{BaFeFoRaWi} provides a connection to various
integrable systems such as Toda field theory \cite{BiGe} and
the generalized KdV hierarchy \cite{Ya}.
In the present case it is natural to expect the super Liouville model
coupled to Wess-Zumino-Novikov-Witten (WZNW) models or the KdV hierarchy
coupled to affine Lie algebras to arise.
In the bosonic case, this kind of integrable system has been partially
studied \cite{BaTjDr}.

One might further generalize the hamiltonian reduction procedure to
any Lie superalgebra. This would then give rise to a super $W$-algebra
coupled to WZNW models.

\newpage
\alphsec
\appe{A. The root system }
In this appendix, we describe the root systems of the Lie
superalgebras  with the even subalgebra $G\oplus A_{1}$ as given in
table \ref{ta:li}.
$\T$ is the simple root of $A_{1}$.
We use the orthonormal basis $e_{i}$ ($i\geq 1$) with positive
metric and $\D_{j}$ ($j\geq 1$) with negative metric:
\eq
e_{i}\cdot e_{j}=\D_{i j}, \quad
\D_{i}\cdot\D_{j}=-\D_{i j}, \quad
e_{i}\cdot \D_{j}=0.
\en
\begin{enumerate}
\item $A(n|1)$ $(n\geq 1)$,
(rank $n+2$,  the dual Coxeter number $h^{\vee}=n-1$) \\
Simple roots:
$\A_{1}=\D_{1}-e_{1}$,   \quad
$\A_{i}=e_{i-1}-e_{i}$, ($i=2, \ldots, n+1$), \quad
$\A_{n+2}=e_{n+1}-\D_{2}$.  \\
Positive even roots:
$e_{i}-e_{j}$, ($1\leq i<j\leq n+1$),  \quad
$\theta=\D_{1}-\D_{2}$. \\
Positive odd roots:
$\D_{1}-e_{j}$,  \quad $e_{j}-\D_{2}$, ($j=1,\ldots, n+1$).
\item $B(n|1)$ $(n\geq 0)$,
 (rank $n+1$,  $h^{\vee}=3-2n$)\\
Simple roots:
$\A_{1}=e_{1}-\D_{1}$, \quad
$\A_{i+1}=\D_{i}-\D_{i+1}$,  ($i=1, \ldots, n-1$), \quad
$\A_{n+1}=\D_{n}$. \\
Positive even roots:
$\theta=2e_{1}$, \quad
$\D_{i}\pm\D_{j}$,   ($1\leq i < j \leq n$),
$\D_{i}$,  ($i=1, \ldots, n$)  \\
Positive odd roots:
$e_{1}$, \quad
$e_{1}\pm\D_{j}$,
($j=1,\ldots, n$).
\item $D(n|1)$ $(n\geq 2)$,
 (rank $n+1$, $h^{\vee}=4-2n$)\\
Simple roots:
$\A_{1}=e_{1}-\D_{1}$,   \quad
$\A_{i+1}=\D_{i}-\D_{i+1}$,  ($i=1, \ldots, n-1$), \quad
$\A_{n+1}=\D_{n-1}+\D_{n}$. \\
Positive even roots:
$\theta=2e_{1}$, \quad
$\D_{i}\pm\D_{j}$,   ($1\leq i < j \leq n$). \\
Positive odd roots:
$e_{1}\pm\D_{j}$, \quad  ($j=1,\ldots, n$).
\item  $D(2|n)$ $(n\geq 1)$
 (rank $n+2$, $h^{\vee}=2n-2$\\
Simple roots:
$\A_{1}=-\D_{2}-\D_{1}$, \quad
$\A_{2}=\D_{1}-e_{1}$, \quad
$\A_{i+2}=e_{i}-e_{i+1}$,  ($i=1, \ldots, n-1$), \quad
$\A_{n+2}=2e_{n}$. \\
Positive even roots:
$-\D_{2}-\D_{1}$,
$\theta=\D_{1}-\D_{2}$, \quad
$e_{i}\pm e_{j}$, ($1\leq i < j \leq n$). \\
Positive odd roots:
$\D_{1}\pm e_{j}$,
$-\D_{2}\pm e_{j}$,
($j=1,\ldots, n$).
\item $D(2| 1;\A)$ ($\A\neq 0, -1, \infty$),
 (rank  3, $h^{\vee}=0$)\\
Simple roots:
$\A_{1}=\half(\sqrt{2\G}\D_{1}+\sqrt{2(1-\G)}\D_{2}+\sqrt{2}e_{3})$,
\quad
$\A_{2}=-\sqrt{2\G}\D_{1}$,  \quad
$\A_{3}=-\sqrt{2(1-\G)}\D_{2}$,
where $\A=\G/(1-\G)$. \\
Positive even roots:
$\A_{2}$, \quad
$\A_{3}$, \quad
$\theta=2\A_{1}+\A_{2}+\A_{3}$. \\
Positive odd roots:
$\A_{1}$, \quad
$\A_{1}+\A_{2}$, \quad
$\A_{1}+\A_{3}$, \quad
$\A_{1}+\A_{2}+\A_{3}$.
\item  $F(4)$
(rank:$4$, $h^{\vee}= -3$)\\
Simple roots:
$\A_{1} = {1\over 2}(\sqrt{3}e_{1}+\D_{1}+\D_{2}+\D_{3})$, \quad
$\A_{2} = -\D_{1}$,  \quad
$\A_{3} = \D_{1}-\D_{2}$,  \quad
$\A_{4} = \D_{2}-\D_{3}$. \\
Positive even roots:
$\A_{2}$,\quad
$\A_{3}$,  \quad
$\A_{4}$,  \quad
$\A_{2}+\A_{3}$, \quad
$\A_{3}+\A_{4}$, \quad
$2\A_{2}+\A_{3}$, \quad
$\A_{2}+\A_{3}+\A_{4}$,
$2\A_{2}+\A_{3}+\A_{4}$, \quad
$2\A_{2}+2\A_{3}+\A_{4}$, \quad
$\theta=2\A_{1}+3\A_{2}+2\A_{3}+\A_{4}$. \\
Positive odd roots:
$\A_{1}$, \quad
$\A_{1}+\A_{2}$, \quad
$\A_{1}+\A_{2}+\A_{3}$, \quad
$\A_{1}+2\A_{2}+\A_{3}$, \quad
$\A_{1}+\A_{2}+\A_{3}+\A_{4}$, \quad
$\A_{1}+2\A_{2}+\A_{3}+\A_{4}$,  \quad
$\A_{1}+2\A_{2}+2\A_{3}+\A_{4}$, \quad
$\A_{1}+3\A_{2}+2\A_{3}+\A_{4}$.
\item $G(3)$
(rank:$3$,  $h^{\vee}=2$)\\
Simple roots:
$\A_{1}=\sqrt{{2\over3}}\D_{1}+{2e_{1}-e_{2}-e_{3}\over 3}$, \quad
$\A_{2}={-e_{1}+2e_{2}-e_{3}\over 3}$,\quad
$\A_{3}=-e_{2}+e_{3}$. \\
Positive even roots:
$\A_{2}$, \
$\A_{3}$, \
$\A_{2}+\A_{3}$, \
$2\A_{2}+\A_{3}$, \
$3\A_{2}+\A_{3}$, \
$3\A_{2}+2\A_{3}$, \
$\theta=2\A_{1}+4\A_{2}+2\A_{3}$. \\
Positive odd roots:
$\A_{1}$, \quad
$\A_{1}+\A_{2}$, \quad
$\A_{1}+\A_{2}+\A_{3}$,\quad
$\A_{1}+2\A_{2}+\A_{3}$, \quad
$\A_{1}+3\A_{2}+\A_{3}$, \quad
$\A_{1}+3\A_{2}+2\A_{3}$, \quad
$\A_{1}+4\A_{2}+2\A_{3}$.
\end{enumerate}

\newpage
\begin{table}
\caption{Lie Superalgebras with an even subalgebra $A_{1}$}
\label{ta:li}
\begin{center}
\begin{tabular}{llllll}                         \hline\hline
$\BG$        & \sdim &rank &$h^{\vee}$  & $\BG_{0}$      & $G$  \\ \hline
$A(n|1)$     & $n^{2}-2n$ ($-2$ for $n=1$) & $n+2$ &$n-1$
& $A_{n}\oplus A_{1}\oplus u(1)$  & $A_{n}\oplus u(1)$  \\
$B(n|1)$     & $(n-1)(2n-1)$ &$n+1$ &$3-2n$
& $B_{n}\oplus A_{1}$  & $B_{n}$ \\
$D(n|1)$     & $(n-1)(2n-3)$ &$n+1$ &$4-2n$
& $D_{n}\oplus A_{1}$  & $D_{n}$ \\
$D(2|n)$     & $(n-2)(2n-3)$ & $n+2$& $2n-2$
& $A_{1}\oplus A_{1}\oplus C_{n}$ & $A_{1}\oplus C_{n}$ \\
$D(2|1;\A)$  & $1$ &$3$ &$0$  & $A_{1}\oplus A_{1}\oplus A_{1}$
& $A_{1}\oplus A_{1}$  \\
$F(4)$       & $8$ &$4$ &$-3$  & $B_{3}\oplus A_{1}$  & $B_{3}$ \\
$G(3)$       & $3$ &$3$ &$2$  & $G_{2}\oplus A_{1}$ & $G_{2}$  \\ \hline
$B(1|n)$     & $(n-1)(2n-3)$ & $n+1$ &$2n-1$
& $A_{1}\oplus C_{n}$  & $C_{n}$ \\ \hline\hline
\end{tabular}
\end{center}
\end{table}

\begin{table}
\caption{Central charges for $G$ extended superconformal algebras}
\label{ta:ce}
\begin{center}
\begin{tabular}{ll}                                 \hline
$\BG$        &    $c_{total}$                          \\ \hline
$A(n|1)$     &    ${6k^{2}+k(n^{2}+3n-9)-n^{2}-2n+3\over k+n-1}$ \\
$B(n|1)$     &    ${(k+1)(4n^{2}+4n-15-6k)\over 2(k+3-2n)}$ \\
$D(n|1)$     &    ${(k+1)(4n^{2}-16-6k)\over 2(k+4-2n)}$  \\
$D(2|n)$     &    ${6 k^{2}+k(2n^{2}+3n-8)+4-4n^{2} \over k+2n-2}$  \\
$D(2|1;\A)$  &    $ -3-6k$ \\
$F(4)$       &    ${2(-2k^{2}+7k+9)\over k-3}$    \\
$G(3)$       &    ${9k^{2}+13k-2\over 2(k+2)}$    \\ \hline
\end{tabular}
\end{center}
\end{table}

\begin{table}
\caption{}
\label{ta:we}
\begin{center}
\begin{tabular}{ll}                                 \hline
$\BG$        & $\bLM^{*}$    \\ \hline
$A(n|1)$     & $(\blm_{1}\oplus\nu), \blm_{n}\oplus(-\nu)$    \\
$B(n|1)$     & $\blm_{1}$    \\
$D(n|1)$     & $\blm_{1}$    \\
$D(2|n)$     & $\blm_{1}(A_{1})\oplus\blm_{1}(C_{n})$    \\
$D(2|1;\A)$  & $\blm_{1}(A_{1})\oplus\blm_{1}(A_{1})$    \\
$F(4)$       & $\blm_{2}(=\blm_{s})$ :\ spin representation of $so(7)$ \\
$G(3)$       & $\blm_{1}$    \\ \hline
\end{tabular}
\end{center}
\end{table}
\end{document}